\documentclass[preprint,12pt,authoryear]{elsarticle}
\usepackage{lineno}
\usepackage{amssymb}
\usepackage{amsmath}
\usepackage{url}
\usepackage{hyperref}
\usepackage{xcolor}
\newcommand{\rev}[1]{\textcolor{black}{#1}}
\newcommand{\revrev}[1]{\textcolor{black}{#1}}

\journal{Neural Networks}

\begin{document}

\begin{frontmatter}



\title{Spurious reconstruction from brain activity} 


\author[inst1,inst2]{Ken Shirakawa} 

\author[inst1,inst2]{Yoshihiro Nagano}

\author[inst1,inst2]{Misato Tanaka}

\author[inst1,inst2]{Shuntaro C. Aoki}

\author[inst1]{Yusuke Muraki}

\author[inst3]{Kei Majima}

\author[inst1,inst2,inst4]{Yukiyasu Kamitani}

\affiliation[inst1]{organization={Graduate School of Informatics, Kyoto University},
            addressline={Yoshida-honmachi, Sakyo-ku}, 
            city={Kyoto},
            postcode={606-8501}, 
            country={Japan}}

\affiliation[inst2]{organization={Computational Neuroscience Laboratories, Advanced Telecommunications Research Institute International},
            addressline={Seika-cho}, 
            city={Sorakugun},
            postcode={619-0288}, 
            country={Japan}}

\affiliation[inst3]{organization={Institute for Quantum Life Science, National Institutes for Quantum Science and Technology},
            city={Chiba},
            postcode={263-8555}, 
            country={Japan}}

\affiliation[inst4]{organization={Guardian Robot Project, RIKEN},
            addressline={Seika-cho}, 
            city={Sorakugun},
            postcode={619-0288}, 
            country={Japan}}

\begin{abstract}
Advances in brain decoding, particularly in visual image reconstruction, have sparked discussions about the societal implications and ethical considerations of neurotechnology. As reconstruction methods aim to recover visual experiences from brain activity and achieve prediction beyond training samples (zero-shot prediction), it is crucial to assess their capabilities and limitations to inform public expectations and regulations. Our case study of recent text-guided reconstruction methods, which leverage a large-scale dataset (Natural Scene Dataset, NSD) and text-to-image diffusion models, reveals critical limitations in their generalizability, demonstrated by poor reconstructions on a different dataset. UMAP visualization of the text features from NSD images shows limited diversity with overlapping semantic and visual clusters between training and test sets. We identify that clustered training samples can lead to ``output dimension collapse,'' restricting predictable output feature dimensions. While diverse training data improves generalization over the entire feature space without requiring exponential scaling, text features alone prove insufficient for mapping to the visual space. Our findings suggest that the apparent realism in current text-guided reconstructions stems from a combination of classification into trained categories and inauthentic image generation (hallucination) through diffusion models, rather than genuine visual reconstruction. We argue that careful selection of datasets and target features, coupled with rigorous evaluation methods, is essential for achieving authentic visual image reconstruction. These insights underscore the importance of grounding interdisciplinary discussions in a thorough understanding of the technology's current capabilities and limitations to ensure responsible development. 
\end{abstract}



\end{frontmatter}


\section{Introduction}
Brain decoding has been widely used in the neuroscience field, revealing 
specific contents of the mind \citep{haxby_distributed_2001, kamitani_decoding_2005, soon_unconscious_2008, horikawa_neural_2013}. 
As brain decoding is sometimes referred to as ``mind-reading'' in popular media 
\citep{somers_science_2021, whang_i_2023, raasch_mind_2023}, it has attracted 
significant attention beyond the scientific community due to its potential for 
real-world applications in medicine and industry. Such neurotechnology has also 
started to affect future ethical discussions and legal regulations 
\citep{unesco_unveiling_2023}. To prevent misleading public expectations and 
policies, scientists need to carefully assess the current status of brain 
decoding techniques and clarify the possibilities and limitations. 

One of the major challenges in brain decoding is the limited amount of brain data 
we can collect. The current brain measurement devices are costly, yielding far 
less brain data than the amounts typically used in image or text processing 
within the field of computer science and AI \citep{deng_imagenet:_2009, schuhmann_laion-5b_2022}. 
Although we have gradually increased the amount of brain data per subject \citep{van_essen_human_2012,
allen_massive_2021,naselaris_extensive_2021,hebart_things-data_2023, xu_alljoined_2024}, 
it remains impractical to collect brain data covering the full range of 
cognitive states and perceptual experiences. The scarcity of brain data limits the applicability and scalability of classification-based decoding approaches, which are primarily developed in the early stage of this field. Such approaches can only decode information confined to the same stimuli or predefined categories used in the training phase, rendering them insufficient for uncovering the neural representation under general or natural conditions.

To overcome this limitation, several decoding methods have been developed to 
enable the prediction of novel contents from brain activities that are not 
encountered during the training phase. \cite{kay_identifying_2008} 
proposed a general visual decoding approach via a statistical encoding model 
that predicted fMRI voxel values from image features. It successfully identified 
novel test images from a set of $1,000$ candidates. \cite{mitchell_predicting_2008} 
utilized co-occurrence rates of specific verb sets for nouns and built a 
computational model to predict fMRI voxel values while thinking about nouns 
in presented line drawing images. Their model demonstrated the ability to 
predict voxel values for novel nouns not seen during the training phase.  
\cite{brouwer_decoding_2009} constructed a color-tuning 
model and predicted brain activity while the subjects were presented with 
color stimuli. As their training stimuli covered most of the color space, 
their methods successfully identified novel colors not included in the 
training dataset. \cite{horikawa_generic_2017} utilized 
deep neural network (DNN) features to decode brain activity measured while 
subjects perceived natural images. They showed successful prediction of 
novel object categories not encountered during the training phase.

In the field of machine learning, ``zero-shot'' prediction refers to the 
ability of a model to accurately predict or classify novel contents not encountered during the training phase \citep{larochelle_zero-data_2008, palatucci_zero-shot_2009}. 
This ability has emerged in various applications across different domains, 
including image classification \citep{radford_learning_2021}, image generation \citep{ramesh_zero-shot_2021}, 
and natural language processing \citep{brown_language_2020}. The concept of 
zero-shot prediction can be considered analogous to brain decoding techniques 
that aim to interpret brain activity patterns associated with previously 
unseen stimuli or experiences. Both approaches seek to generalize knowledge 
gained from a limited set of training data to novel situations, enabling the 
interpretation of new information without explicit prior exposure. To achieve 
effective zero-shot prediction, the model often utilizes a compositional 
representation of the output \citep{lake_building_2017, 
higgins_towards_2018}. 
Compositional representation enables the understanding and generating of novel features through the combination of previously learned ones. 
By learning the underlying structure and relationships between different 
\rev{features}, the model can generalize its knowledge to new, unseen instances. 

Visual image reconstruction is another prominent example of zero-shot 
prediction in brain decoding. This task aims to recover perceived novel images 
that were not encountered during the training phase, effectively reconstructing 
visual experiences from brain activity patterns \citep{stanley_reconstruction_1999,
miyawaki_visual_2008}. As our perceptual visual experiences cannot be 
fully covered by limited brain data, reconstruction methods require strong 
generalizability. \cite{miyawaki_visual_2008} conducted a study 
demonstrating the reconstruction of perceived arbitrary $10 \times 10$ binary-contrast images from brain activity. They built multiple modular decoders to predict 
the local contrasts of each location and combined their predictions. This 
approach leverages the compositional representation of the visual field, 
which is organized retinotopically in the early visual cortex. Incorporating 
cortical organization into the model's architecture can improve 
its ability to perform zero-shot prediction and reconstruct novel visual 
experiences from brain activity. Although the training stimuli were only 
$400$ random images, it was possible to reconstruct an arbitrary image from a 
set of possible $2^{100}$ instances, including geometric shapes such as crosses 
and alphabets. Similarly, \cite{shen_deep_2019} replaced local 
decoders with DNN feature decoders. Although their training stimuli were 
$1,200$ natural images, they demonstrated reconstructing novel images, 
including artificial images, which were not part of the training set. 
These successes suggest that the proposed reconstruction models capture 
rich and comprehensive information about the general aspects of the neural 
representation, beyond merely the information defined by the training data \citep{kriegeskorte_interpreting_2019}. 
Developing reliable reconstruction methods also enables further analysis of 
subjective visual experiences, such as visual imagery \citep{shen_deep_2019}, 
attention \citep{horikawa_attention_2022}, and illusion \citep{cheng_reconstructing_2023}. 
Decoding novel brain states that were never encountered 
during the training phase can be a promising approach to neural mind-reading 
\citep{kamitani_decoding_2005}.

\begin{figure}[!t]
  \centering
    \includegraphics[width=1.0\linewidth]{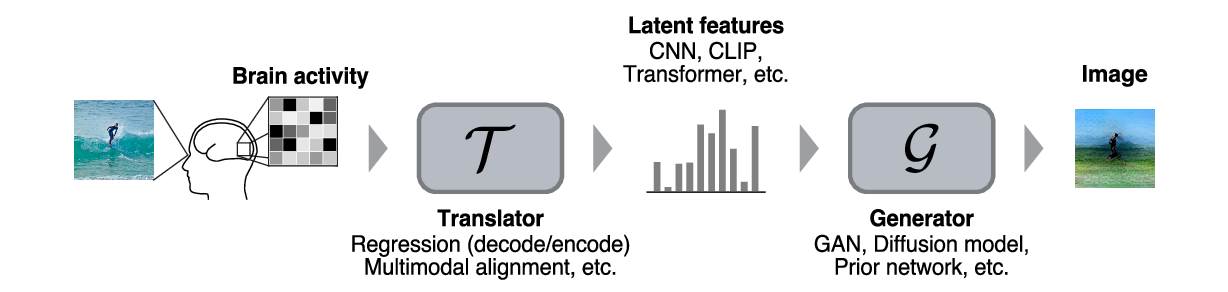}
  \caption{{\textbf {Visual image reconstruction pipeline. }}
  The first step involves translating brain activity patterns into machine/AI latent representations. While this process typically involves brain decoding of latent features using machine learning, here we term it ``translation.'' This terminology acknowledges that brain activity itself can be viewed as a latent representation of an image, framing the process as a translation from neural to machine latent representations. In the second step, a generator module takes these translated latent features as input and converts them into a visual image that corresponds to the content represented by the original brain activity.}
  \label{Figure_01}
  \end{figure}

Visual image reconstruction pipelines typically comprise three main components: translator, latent features, and generator  (Fig.~\ref{Figure_01}). The translator converts brain activity patterns into a latent feature space, employing either linear regression \citep{shen_deep_2019,seeliger_generative_2018,
 mozafari_reconstructing_2020,ozcelik_reconstruction_2022} or nonlinear transformation \citep{qiao_biggan-based_2020}. Latent features serve as surrogate representations of perceived visual images, evolving from primitive forms like local contrasts \citep{miyawaki_visual_2008} to more sophisticated DNN features, such as intermediate outputs of recognition models \citep{horikawa_generic_2017, shen_deep_2019}. In our current work, we reframe this process as ``translation'' rather than ``feature decoding,'' a term we used in previous studies. This terminology acknowledges two important points: first, brain activity itself can be considered a latent representation of an image or the perception formed by it and second, it helps avoid potential ambiguity between image encoding/decoding and brain encoding/decoding processes. This new perspective conceptualizes the process as a translation between two latent spaces: from neural representation to machine representation.

The generator visualizes translated latent features into images. Some studies have used pretraining image generative models for the generator module \citep{mozafari_reconstructing_2020,
 qiao_biggan-based_2020,ozcelik_reconstruction_2022}. Image optimization can also be regarded as a generator \citep{shen_deep_2019}. End-to-end mapping from brain activity to images using DNNs can also be considered to contain these components implicitly or as a generator-only method \citep{Fujiwara2013Modular, shen_end_end_2019,
 beliy_voxels_2019, ren_reconstructing_2021, 
 gaziv_self-supervised_2022,lin_mind_2022, 
 chen_seeing_2023}. 

Recent advances in generative AI, particularly in text-to-image generation, 
have naturally given rise to expectations that these techniques could provide 
a valuable tool for visual image reconstruction by leveraging semantic representations. 
In addition, there has been a growing trend towards collecting neural datasets 
using a wide range of diverse visual and semantic content. This shift aims to 
capture a more comprehensive and ecologically valid representation of the human 
experience \citep{naselaris_extensive_2021}. Researchers have started to collect 
large-scale fMRI datasets, such as the Natural Scene Dataset (NSD; \citealp{allen_massive_2021}) 
and the THINGS-fMRI dataset \citep{hebart_things-data_2023}, which include more than $10,000$ brain samples per subject. These datasets incorporate a broader range of brain data induced by diverse visual stimuli with text or 
category annotations. Importantly, recent studies have demonstrated that combining large-scale datasets with generative AI techniques can lead to more realistic reconstructions from brain activity \citep{takagi_high-resolution_2023, ozcelik_natural_2023,scotti_reconstructing_2023, bai_dreamdiffusion_2024, benchetrit_brain_2024, scotti_mindeye2_2024}. \rev{These approaches commonly utilize linear regression models as translators, contrastive language-image pretraining (CLIP) text features \citep{radford_learning_2021} as part of the latent features, and text-to-image diffusion models \citep{ramesh_zero-shot_2021,
xu_versatile_2022} as generators. MindEye2 \citep{scotti_mindeye2_2024} has recently shown improved reconstruction performance on the NSD, using a nonlinear translator, latent features of a CLIP’s image model, and a fine-tuned generator. This approach also includes a refinement step that enhances the realism of reconstructed images.
}

While these recent approaches show promising results, it remains uncertain 
whether these methods truly achieve zero-shot reconstruction due to several 
factors. The complex model architectures employed in these studies, along 
with the use of a large-scale dataset, make it challenging to interpret and 
understand the underlying mechanisms driving the reconstruction process. 
To fully assess the zero-shot prediction capabilities of these approaches, 
it is essential to rigorously test their generalizability across different 
datasets and to provide detailed analyses of the individual model components. 
This test includes evaluating the performance of the translators, latent features, 
and generators used in these methods. Furthermore, the characterization of 
the diversity of stimuli in the datasets and the latent representations has not 
been thoroughly explored. It is unclear whether the recently proposed 
datasets, such as the NSD, are optimally designed to capture the full range 
of human visual experiences and to support the development of truly 
generalizable prediction models. 

In the following, we begin with a case study that critically tests text-guided reconstruction methods. We evaluate the approaches of \cite{takagi_high-resolution_2023} and \cite{ozcelik_natural_2023}, which were originally evaluated using the NSD (\rev{additionally, \cite{scotti_mindeye2_2024}'s method is examined}). Our analysis reveals limitations in these methods. First, reconstruction quality substantially degrades when tested on a dataset specially designed to avoid object category overlaps between training and test sets \citep{shen_deep_2019}. Second, the post-hoc selection procedure used by \cite{takagi_high-resolution_2023} can produce seemingly convincing reconstructions even from random brain data when applied to the NSD. Further investigation reveals limited semantic and visual diversity in the NSD stimulus set, with few distinct semantic clusters, potentially explaining these issues. We also demonstrate the failure of zero-shot prediction in the latent feature space and the inability to recover a stimulus from its latent features, suggesting fundamental constraints in the latent feature representation. These findings indicate that the apparent realism of reconstructions likely results from classification into clusters shared between training and test sets, combined with hallucinations by the generative model.

In the formal analysis and simulation section, we investigate the general factors underlying the limitations observed in our case study. We introduce the phenomenon of ``output dimension collapse'' that occurs when translating brain activity into latent feature space. Our analysis shows that regression models trained on clustered targets become overly specialized to training examples, causing their outputs to collapse into a restricted subspace of the training set. Through systematic simulations with clustered data, we demonstrate that successful out-of-sample prediction requires the number of training clusters to scale linearly with feature dimensionality, suggesting that zero-shot prediction becomes feasible given sufficient stimulus diversity and compositional representations. We also discuss the caveats associated with evaluating reconstructions using identification metrics alone and explore the preservation of image information at hierarchical layers of DNNs. Finally, we provide general accounts on how we could be fooled by seemingly realistic reconstructions generated by AI models. Based on these analyses, we conclude with recommendations for developing more reliable reconstruction methods and establishing rigorous evaluation protocols.

\section{Results}
\subsection{Case study}
We \rev{primarily} investigated two recent generative AI-based reconstruction methods, 
StableDiffusionReconstruction \citep{takagi_high-resolution_2023} and 
Brain-Diffuser \citep{ozcelik_natural_2023}, as well as their validation dataset, 
the Natural Scene Dataset (NSD; \citealp{allen_massive_2021}). \rev{ We selected these two methods for three key reasons. First, these methods represent reconstruction approaches that have gained significant public attention by leveraging recent advances in generative AI. Second, both utilize the NSD, which currently serves as a widely adopted benchmark for predictive modeling in the field. Third, these methods employ straightforward linear translators through ridge regression, an approach that has become standard practice}. Both reconstruction methods utilize 
CLIP features \citep{radford_learning_2021} to effectively apply recent 
text-to-image diffusion models in visual image reconstruction analysis. 
CLIP text features are obtained from the average of five text annotations 
corresponding to the stimulus image. This text annotation information is 
only used during training to map the brain activity into the CLIP text features. 
In the test phase, they directly translate CLIP text features from brain activity 
during image perception. Hereafter, these two reconstruction methods will 
be referred to together as text-guided reconstruction methods. \rev{We also replicated the MindEye2 reconstruction method \citep{scotti_mindeye2_2024}, a more recent study evaluated using the NSD. Unlike the other two methods, MindEye2 uses a variant of the CLIP model's image embeddings as latent features, implements a nonlinear translator, and utilizes specially designed diffusion-based generators rather than relying on text features as direct decoding targets.}

The StableDiffusionReconstruction method \citep{takagi_high-resolution_2023} uses components of the Stable Diffusion model \rev{\citep{rombach_high-resolution_2022}}, 
the VAE features \citep{kingma_auto-encoding_2014}, and CLIP text features 
as latent features. Similarly, the Brain-Diffuser method \citep{ozcelik_natural_2023} 
utilizes components of another type of diffusion model \citep{xu_versatile_2022}, 
CLIP text features, CLIP vision features, and VDVAE features \citep{child_very_2021} 
as latent features. Both methods translate brain activity into their latent features \rev{using a linear translator, and initial images are first generated from translated VAE/VDVAE features.} These initial images are then passed through the 
image-to-image pipeline of the diffusion model conditioned on the translated 
CLIP features, producing the final reconstructed images. They validated the 
reconstruction performance using the NSD dataset, preparing training and test 
data based on the data split provided by the NSD study. Thanks to the authors' efforts in making the datasets and scripts publicly 
available, we were able to conduct our replication analysis effectively. 
We compared the reconstructed results of these two text-guided reconstruction 
methods (additionally, MindEye2) with those from a previous image reconstruction method, iCNN \citep{shen_deep_2019}. 
For more information on datasets and reconstruction methods, refer to Methods (``Datasets'' and ``Reconstruction methods'') or the original studies 
\citep{allen_massive_2021,
shen_deep_2019,takagi_high-resolution_2023,ozcelik_natural_2023}.

\subsubsection{Observations: Failed replication and convincing reconstruction from random data} 

\begin{figure}[!t]
  \centering
    \includegraphics[width=1.0\linewidth]{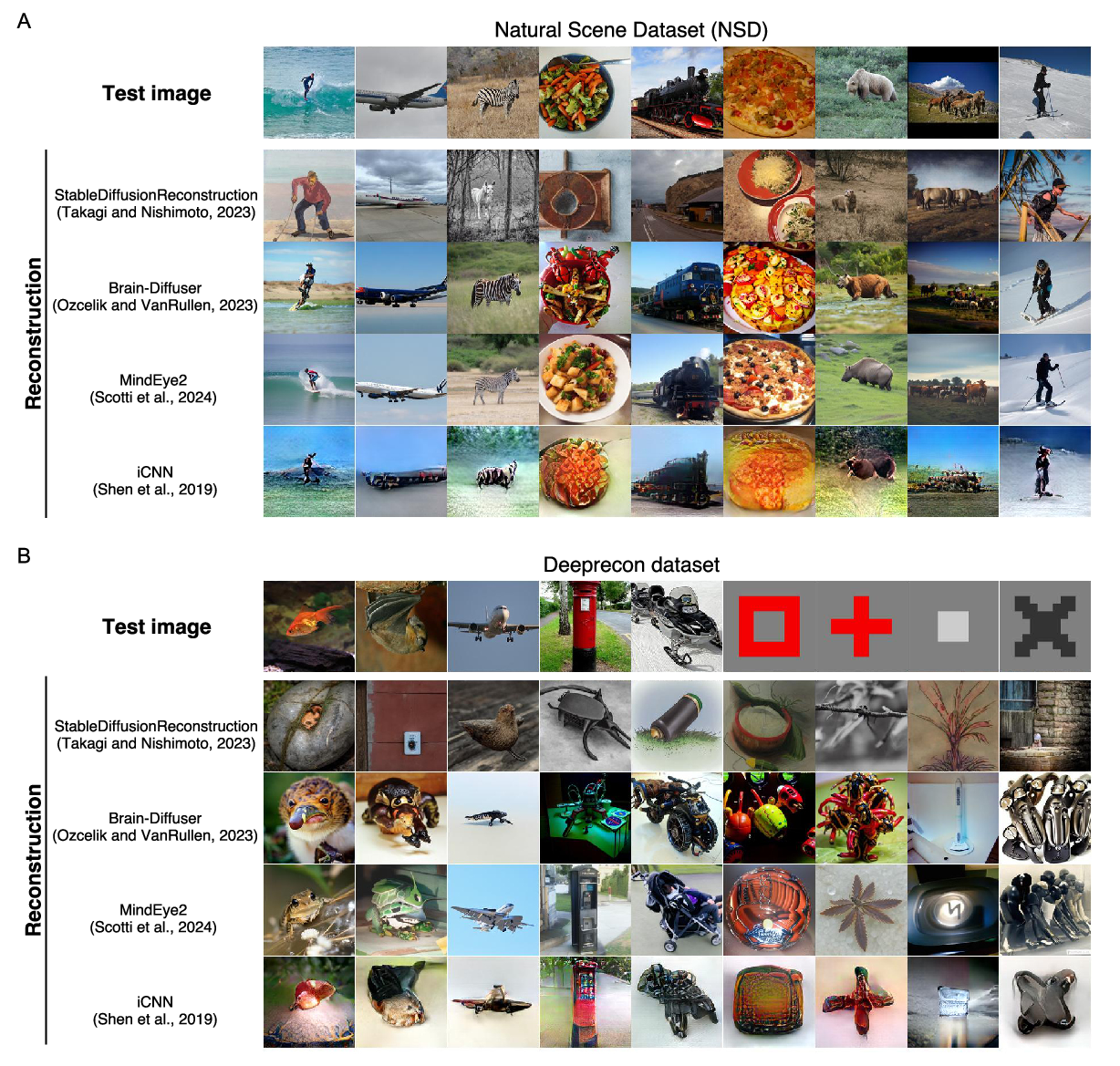}
  \caption{{\textbf {Comparison of image reconstruction results across datasets and methods.} }
  (\textbf{A}) Reconstruction results for the Natural Scene Dataset (NSD). The first 
  row shows the original test images, followed by reconstructions using 
  StableDiffusionReconstruction, Brain-Diffuser, MindEye2, and iCNN methods. 
  (\textbf{B}) Reconstruction results for the Deeprecon dataset, presented in the same 
  format as (\textbf{A}).
   }
  \label{Figure_02}
\end{figure}

We first confirmed the reproduction of the findings of the original methods (Fig.~\ref{Figure_02}A).  The reconstructed images produced by the StableDiffusionReconstruction method \citep{takagi_high-resolution_2023} showed slightly degraded performance compared to the original paper, but still successfully captured the semantics of the test images.  The Brain-Diffuser method \citep{ozcelik_natural_2023} effectively captured most of the layout and semantics of the test images when applied to the NSD dataset. Similarly, MindEye2 \citep{scotti_mindeye2_2024} generated reconstructions that preserved key visual elements of the original stimuli. Notably, despite originally being validated on a different dataset (Deeprecon), the iCNN method \citep{shen_deep_2019} also performed well on the NSD dataset, capturing the dominant structures of the objects. This performance is consistent with the findings reported in the original study.

To further investigate the generalizability of the text-guided reconstruction 
methods, we attempted to replicate their performance using a different 
dataset, Deeprecon, which was originally collected for the study by \cite{shen_deep_2019}. The Deeprecon dataset was explicitly 
designed to avoid overlap between training and test sets, making it a suitable benchmark for evaluating the zero-shot prediction capabilities of reconstruction 
methods. However, the original Deeprecon dataset lacked the text annotations 
required by the text-guided reconstruction methods. To enable a fair 
comparison, we collected five text annotations for each training stimulus in 
the Deeprecon dataset through crowd-sourcing and used them to generate CLIP 
text features.

Despite our careful replication efforts, including preparing text annotations, both text-guided reconstruction methods and MindEye2 failed to achieve the same level of performance on the Deeprecon dataset as they did on the NSD dataset (Fig.~\ref{Figure_02}B). The reconstructed 
images produced by the text-guided methods exhibited realistic appearances 
but suffered from largely degraded quality compared to their performance on 
the NSD. Notably, the text-guided 
reconstruction methods generated realistic reconstructions even for simple 
geometric shapes in the Deeprecon dataset, which deviated 
strikingly from the original stimuli. \rev{Similarly, the reconstructions from MindEye2 did not resemble the test images but tended to exhibit object categories in the Deeprecon training set (\textit{frog, fighter aircraft, baby buggy, or baseball glove}). Sample size matching between NSD and Deeprecon yielded similar reconstruction quality (Fig.~\ref{Figure_A1}), suggesting that sample size alone does not account for the poorer results on the Deeprecon.} In contrast, the iCNN method consistently 
provided faithful reconstructions for both the NSD and Deeprecon datasets, 
despite its simplicity compared to these methods. These results suggest that the text-guided reconstruction methods and MindEye2 struggle to generalize across different datasets. The tendency to generate realistic yet inaccurate reconstructions, especially for simple shapes, indicates that these methods might rely more on learned training stimuli than on actual brain activity information.

\begin{figure}[!t]
  \centering
    \includegraphics[width=1.0\linewidth]{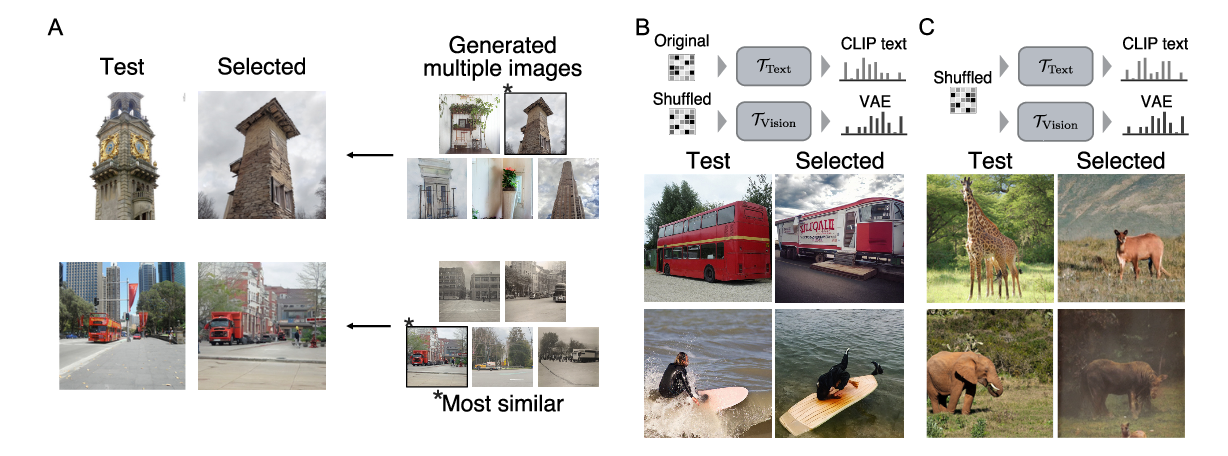}
  \caption{{\textbf {Analysis of post-hoc selection in the StableDiffusionReconstruction method \citep{takagi_high-resolution_2023}.} }
  (\textbf{A}) Selection procedure. Five images were generated, and the one most 
  closely resembling the test image was selected based on perceptual similarity 
  metrics (PSM). (\textbf{B}) Images were generated from the CLIP text features 
  translated from original brain activity and the VAE (vision) features 
  from shuffled brain activity. Examples selected from five generations are 
  shown with the test images. (\textbf{C}) Images were generated from the CLIP text 
  and the VAE (vision) features, both translated from shuffled brain activity. 
  Examples of selected images are shown with the text images. Examples are 
  shown in (\textbf{B}).}

  \label{Figure_03}
\end{figure}

Upon further investigation, we noted a questionable post-hoc image selection 
procedure. In the StableDiffusionReconstruction study \citep{takagi_high-resolution_2023}, 
they presented the reconstruction results by the following procedure: 
``We generated five images for each test image and selected the generated images 
with highest PSM.'' In their paper, PSM refers to perceptual similarity metric, which was calculated from early, middle, and late layers of several image recognition DNNs. This procedure is illustrated in Fig.~\ref{Figure_03}A. This selection might lead readers or peer reviewers, particularly those not specialized in the brain decoding field, 
to overestimate the effectiveness of the methods and potentially lead to 
a distorted understanding of the actual reconstruction performance. 
Note that the Brain-Diffuser and MindEye2 studies did not execute 
such procedures, and in their subsequent report \citep{takagi2023improving}, 
they updated the image presentation procedure more fairly as: ``we generated five images with different stochastic noise and selected three 
images randomly.''

To examine the impact of this post-hoc selection procedure, we conducted an 
experiment using random brain data. Instead of feeding the test brain data to 
trained translators, we shuffled the activities within each voxel of the 
NSD test set independently to create random brain data. Specifically, 
the brain data of the NSD test set is in a matrix shape, with rows representing 
stimulus samples and columns representing voxels. To generate the random 
brain data, we selected one column (voxel) of the matrix and randomly 
shuffled its values. This process was repeated for all voxels independently. 
Surprisingly, when the random brain data were input into the VAE feature 
translator, which contributes to producing initial images, plausible images 
were obtained by generating five images and selecting the best one (Fig.~\ref{Figure_03}B). 
Even more strikingly, when the random brain data were input into both VAE and CLIP text 
feature translators, we still obtained convincing results by simply generating 
images five times and conducting the selection mentioned above (Fig.~\ref{Figure_03}C). 
These observations are inexplicable because the artificially created brain data 
should completely lack any information related to the original visual stimuli.

These observations raise perplexing questions about the performance and 
generalizability of recent reconstruction methods. Their performances largely deteriorated when we switched the evaluated dataset from the NSD to Deeprecon, highlighting the need to better understand why these methods succeed with the NSD.  Moreover, 
the ability to obtain plausible reconstructions from random brain data by 
merely generating multiple images and selecting the best ones suggests that 
there may be fundamental issues with both the evaluation dataset and the 
components of the reconstruction methods themselves. In the following sections, 
we will thoroughly investigate the potential problems associated 
with the NSD dataset and each component of the text-guided reconstruction 
pipeline. 

\subsubsection{Lack of diversity in the stimulus set} 

\begin{figure}[!t]
  \centering
    \includegraphics[width=1.0\linewidth]{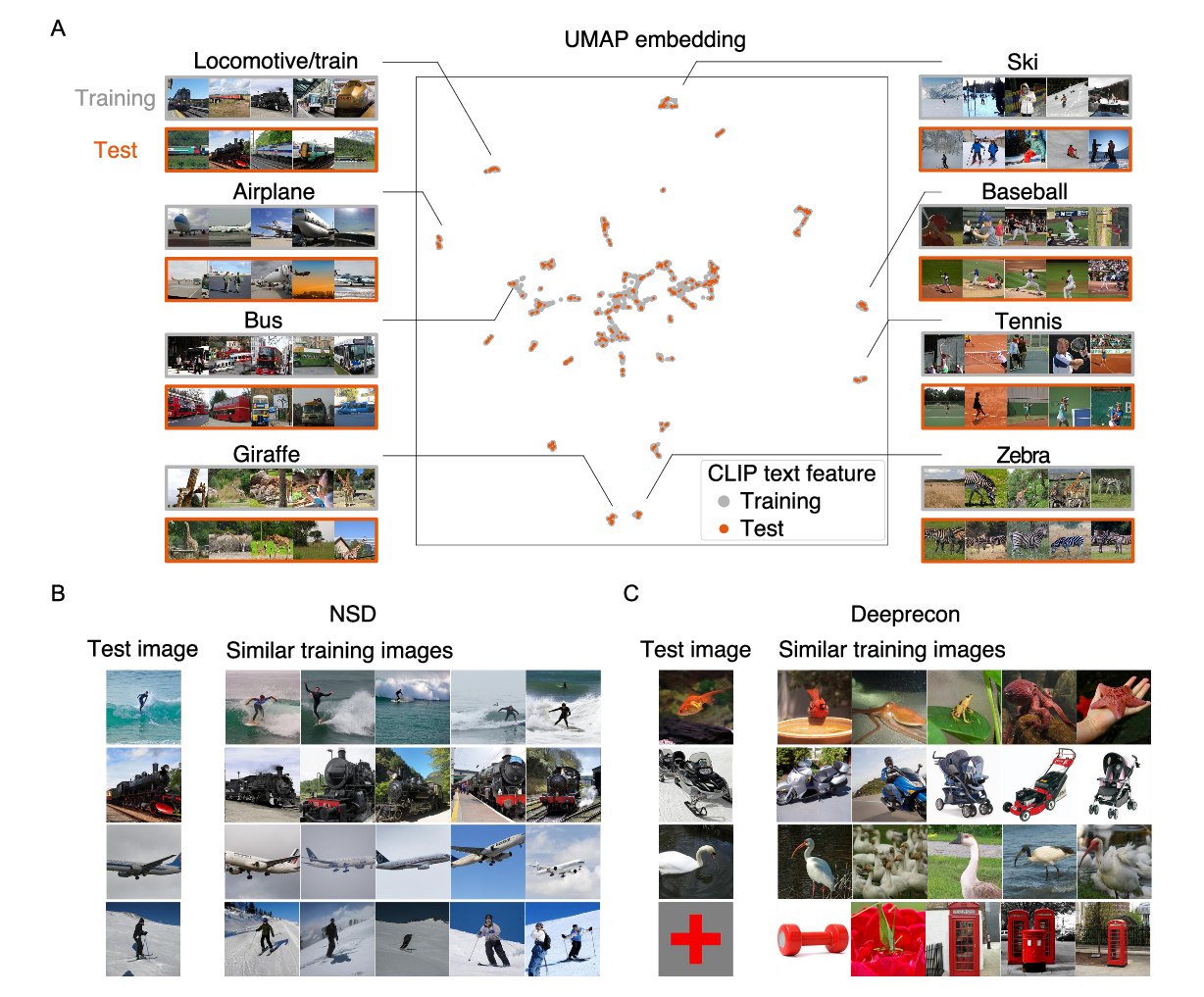}
  \caption{{\textbf {Dataset diversity and similarity between training and test stimuli.}} 
  (\textbf{A}) UMAP visualization of CLIP text features of the NSD stimuli. The center figure shows 
  the scatter plot of the UMAP embedding of CLIP text features. The gray points 
  represent training samples, while the orange points represent test samples. 
  The surrounding images were randomly selected from each cluster. (\textbf{B}, \textbf{C}) 
  Similarity between training and test images.  For each example test image 
  (the one on the left of each row), the five training images with the highest 
  similarity were selected using the DreamSim metric and displayed. This 
  analysis was performed for the NSD (\textbf{B}) and Deeprecon (\textbf{C}) datasets, 
  using the same procedure.
   }
  \label{Figure_04}
\end{figure}

First, we examined the characteristics and limitations of the NSD dataset 
itself. To characterize the diversity of stimuli in the datasets and their 
latent representations, we focused on the CLIP text features, which are used as 
latent features in text-guided reconstruction methods. We employed uniform manifold 
approximation and projection (UMAP) \citep{mcinnes_umap:_2018} to visualize the 
CLIP text features of the NSD stimuli (see Methods ``UMAP visualization''). 
The visualization revealed approximately $40$ distinct clusters, with 
considerable overlap between the training and test sets (Fig.~\ref{Figure_04}A). 
Interestingly, we were able to describe the stimulus images in each cluster 
using a single semantic label, such as \textit{airplane, giraffe, or tennis}. Despite 
the NSD containing around $30,000$ brain samples per subject, the diversity 
of the presented stimuli was quite limited to just around $40$ semantic categories. 
 Here, we performed UMAP visualization using the parameters recommended in the official guide for clustering. Even with the default UMAP parameters, a similar cluster structure was observed (Fig.~\ref{Figure_A2}).
In contrast, the Deeprecon dataset, which was specifically designed to 
differentiate object categories between training and test data, exhibited 
less overlap between the two sets (Fig.~\ref{Figure_A3}; \rev{see Fig.~\ref{Figure_A4} for the latent features used in MindEye2}).

To further investigate the similarity between the training and test stimuli, we analyzed stimulus images using DreamSim, a state-of-the-art perceptual similarity metric \citep{fu_dreamsim_2023}. DreamSim was used to identify the most perceptually similar training images for each test image. We found that the training images identified by DreamSim metric were highly similar to the test images in the NSD, not only in semantic labels but also in overall layout and visual composition (Fig.~\ref{Figure_04}B). In contrast, the same analysis on the Deeprecon dataset revealed that its training images were substantially different from the test images (Fig.~\ref{Figure_04}C). \rev{To further assess whether the similarity between NSD training and test images is excessively high, we measured the similarity between the two sets. As a reference, we also measured the similarity between NSD training images and a large-scale independent dataset (CC3M; \citeauthor{sharma_conceptual_2018}, \citeyear{sharma_conceptual_2018}). This analysis revealed that the NSD test set contained stimuli that were highly similar to the training set, compared to those in the independent dataset (Fig.~\ref{Figure_A5}). In contrast, the Deeprecon test set, due to its carefully designed training--test split, exhibited a similar level of similarity to the independent dataset.}

These findings suggest that the distribution of NSD test 
images is heavily biased toward that of the training images, with 
a significant overlap in the visual and semantic features present in both sets. 
Such a strong bias raises concerns about the actual reconstruction performance 
of methods evaluated on this dataset. The convincing reconstruction results with the NSD may be largely attributed to the methods' tendency to replicate specific characteristics observed in the training set, potentially at the expense of generalizing to novel stimuli. The distinct differences between the NSD and Deeprecon datasets in terms of stimulus similarity highlight the importance of carefully designing evaluation benchmarks to rigorously evaluate the generalization capabilities of visual image reconstruction studies.

\subsubsection{Failed zero-shot prediction in the feature space} 

\begin{figure}[!t]
  \centering
    \includegraphics[width=1.0\linewidth]{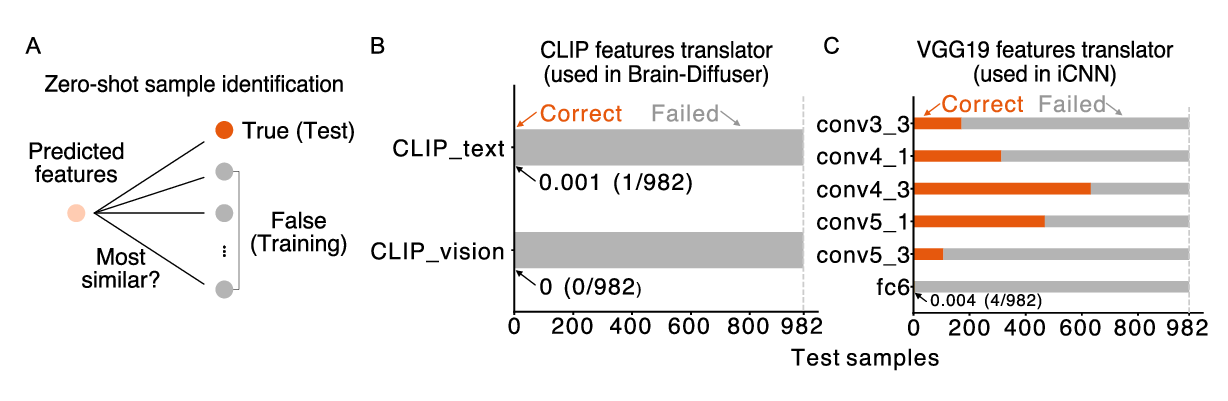}
  \caption{{\textbf {Evaluation of zero-shot sample identification performance.}} 
  (\textbf{A}) Schematic diagram of novel sample identification. Similarity is calculated 
  between predicted (translated) features and test features, as well as between 
  predicted features and each of the training features. Identification accuracy 
  is determined by how often the predicted features show the highest similarity 
  to the true test features compared to all training features. (\textbf{B}) Zero-shot 
  identification performance using predicted CLIP features (text and vision) 
  from the Brain-Diffuser method, evaluated on the NSD dataset. (\textbf{C}) Zero-shot 
  identification performance using predicted features from several intermediate layers of 
  the VGG19 model employed in the iCNN method, also evaluated on the NSD 
  dataset. For both (\textbf{B}) and (\textbf{C}), the identification task involved $8,859$ 
  training samples plus one test sample, resulting in an $8,860$-way identification. 
  The chance performance level is thus $1/8860$. 
   }
  \label{Figure_05}
\end{figure}

Given that many of the NSD test images closely resemble those in the training set, \rev{it is uncertain whether the translator’s predictions genuinely capture new, unseen stimuli (\textit{i.e.}, zero-shot generalization) or simply replicate features of training images that share similar semantics and visual structure. }To probe this, we evaluated the zero-shot prediction capability of the CLIP feature translators by conducting an $(N+1)$-way identification analysis, where $N$ represents the entire training set ($8,859$ samples for subject 1 in the NSD) and $1$ is the target test sample (Fig.~\ref{Figure_05}A). \rev{Concretely, we measured the correlation between each translator’s predicted features and the true features of (1) the correct test sample and (2) every training sample. We then asked whether the predicted features were most similar to the \rev{true test} features, thereby correctly identifying them. Identification performance above chance suggests that the translator captures information specific to the test sample beyond the learned training patterns, supporting zero-shot prediction. Conversely, poor performance indicates that the translator does not predict the unique properties of the test sample.}

The results show that the identification performance of CLIP feature translators was nearly $0$\% (Fig.~\ref{Figure_05}B). This poor performance suggests that the CLIP feature translator captures only rough semantic categories encountered in the training set rather than fine-grained, instance-level details. This finding raises a question about the effectiveness of CLIP features for zero-shot prediction tasks in brain decoding. 

By contrast, the VGG19 features \citep{simonyan_very_2014}, as used in the iCNN method, demonstrated moderate identification performance at the intermediate layers (Fig.~\ref{Figure_05}C). \rev{This success could be attributed to the compositional representation of VGG19’s intermediate features. Unlike CLIP features, which are optimized for semantic alignment across vision and language, VGG19’s intermediate features, extracted through convolutional layers, contain primarily visual representations that retain sufficient local structure. This characteristic allows the translator to generalize to novel images by combining learned local spatial features rather than relying solely on semantic similarity. Consequently, VGG19’s compositional features may help achieve a certain degree of zero-shot prediction, distinguishing new images from closely resembling training examples.
}

\begin{figure}[!p]
  \centering
    \includegraphics[width=1.0\linewidth]{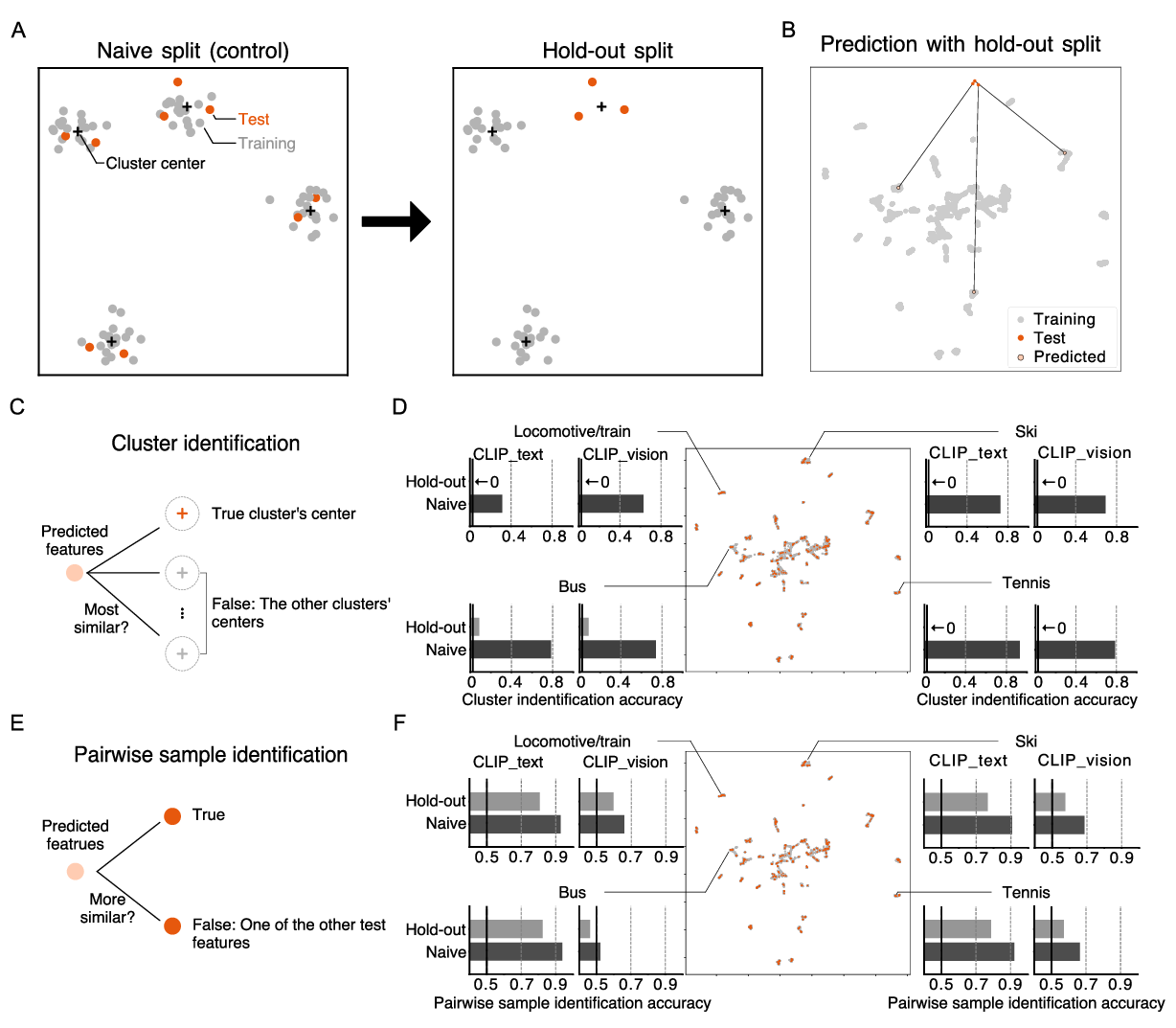}
  \caption{{\textbf {Cluster hold-out analysis.}} 
  (\textbf{A}) Hold-out procedure. For each test sample from a given cluster, all training samples in that cluster are excluded. The naive split (control) uses an equal number of training samples but allows overlap between training and test clusters. (\textbf{B}) 
  Prediction examples. Gray and dark orange points denote training and test samples in the hold-out condition, respectively; light orange points are the predicted latent features, with black lines connecting true features to their predictions. (\textbf{C}) Cluster 
  identification procedure. The predicted features are compared with the cluster centers (average features in each cluster), and the most similar cluster is selected. 
  (\textbf{D}) Cluster identification results. Surrounding panels display CLIP 
  text and vision feature performance in the hold-out and naive split 
  conditions of the four representative semantic clusters. The chance level is $1/40$. 
  (\textbf{E}) Pairwise sample identification procedure. Predicted features are compared with the true test features and with one of the other test features. The sample with features more similar to the prediction is selected. This procedure is repeated for all other test samples, and the proportion of correctly identified true test features is calculated. (\textbf{F}) Pairwise 
  sample identification results. Results are presented similarly to cluster identification, with a chance level of $1/2$.
  }
  \label{Figure_06}
\end{figure}

Furthermore, we investigated whether the CLIP feature translator enables the 
prediction of novel semantic clusters not in the training. We redesigned the 
dataset split to ensure no semantic clusters were shared between them \rev{as 
in previous zero-shot prediction studies} \citep{mitchell_predicting_2008, brouwer_decoding_2009}. 
We first applied $k$-means clustering to the UMAP embedding space of the 
NSD's CLIP text features (Fig.~\ref{Figure_A6}A). We set the number of clusters as
$40$ based on our visual inspection of the UMAP results. Based on these clustering 
results, we performed a hold-out analysis: when predicting samples within a 
cluster (\textit{e.g.}, \textit{ski} cluster), we excluded samples of that cluster from the training set (Fig.~\ref{Figure_06}A right; hold-out split condition). As a control, 
we also prepared a naive data split condition where the training sample size 
is the same as in the hold-out split condition but allows overlapping semantic clusters (Fig.~\ref{Figure_06}A left; naive split condition). When we visualized 
the properties of predicted features in hold-out analysis by transforming them into the previous UMAP embedding space (Fig.~\ref{Figure_04}A), we observed 
that the predicted features tend to diverge largely from their original 
clusters and move into other clusters (Fig.~\ref{Figure_06}B).

To quantitatively assess the performance of the feature translator, we employed 
two identification metrics. The first metric is cluster identification accuracy. 
Cluster identification accuracy focuses on evaluating the translator's ability 
to predict features that correctly identify the semantic cluster to which a 
test sample belongs (Fig.~\ref{Figure_06}C). In this analysis, we calculated the similarity between the predicted features of test samples and the average features of the training samples within each semantic cluster. The accuracy is then calculated as the 
percentage of predicted features that successfully identify the original 
semantic cluster of their corresponding test samples in the latent feature space. 

The cluster identification accuracy in the hold-out split condition exhibited 
a substantial drop compared to the naive split condition across all semantic 
clusters (Fig.~\ref{Figure_06}D). Notably, the cluster identification accuracy 
was frequently $0$\% in the hold-out split condition (see Fig.~\ref{Figure_A6}B for all cluster results). These results expose a severe limitation of the translator when dealing with novel semantic clusters absent from the training set. This suggests that the CLIP feature translator primarily functions as a ``classifier;'' its prediction (translation) heavily relies on predefined semantic features used in a training phase rather than generalizing to new semantic categories. 

The second metric is pairwise sample identification accuracy, a commonly used metric in the evaluation of feature 
prediction and reconstruction performance \citep{beliy_voxels_2019, shen_end_end_2019,
shen_deep_2019, mozafari_reconstructing_2020,qiao_biggan-based_2020,
ren_reconstructing_2021,gaziv_self-supervised_2022,takagi_high-resolution_2023,
ozcelik_natural_2023,scotti_reconstructing_2023,
denk_brain2music_2023,koide-majima_mental_2024}. 
This analysis assesses \rev{whether the translated features for a given test sample are more similar to its actual features than to those of a randomly chosen sample in the test set (Fig.~\ref{Figure_06}E). The accuracy is calculated as the average winning rate of the predicted features against all the test samples, reflecting how often the predicted features are closer to the correct sample than to any alternative.}

Intriguingly, even though the feature translator completely failed to identify the true cluster in the hold-out split condition, pairwise sample identification accuracy often exceeded chance across clusters even in the hold-out split condition (Fig.~\ref{Figure_06}F). \rev{This discrepancy arises because pairwise identification is based on relative similarity: if the translated features are ``less wrong'' for the true sample than for another test sample, they will still be deemed a match. As a result, merely success by a smaller margin than the alternatives can inflate the overall identification score, giving a misleading impression of the translator’s ability to capture new clusters. This observation suggests that relying on pairwise identification accuracy alone, a common practice in many studies, may overestimate reconstruction performance. This issue is further discussed in ``Caveat with evaluation by pairwise identification''}.

\rev{
When applying the hold-out split procedure to the full reconstruction pipeline, StableDiffusionReconstruction showed noticeable degradation, whereas Brain-Diffuser remained more robust (Fig.~\ref{Figure_A7}). This robustness arises because, even in the hold-out split, the training and test stimuli remain highly similar in the NSD (Fig.~\ref{Figure_A7}A). As a result, Brain-Diffuser reconstructions still capture the general visual layout of the original images, with only minor semantic differences (Fig.~\ref{Figure_A7}C: 1-cluster hold-out). However, when removing additional clusters such that the similarity between training and test samples approximates that of a completely independent dataset (Fig.~\ref{Figure_A7}A: similarity-matched hold-out), the reconstructions, though still realistic, deviate considerably from the original images, both semantically and visually (Fig.~\ref{Figure_A7}C: similarity-matched hold-out). This observation suggests that training–test similarity in the NSD played a crucial role in driving text-guided reconstruction performance. When this similarity is reduced to the level of an independent dataset, the ability to reconstruct meaningful details of unseen images diminishes, confirming that the performance of these methods relies on such overlap.
}

\subsubsection{Failed recovery of a stimulus from its latent features} 

Finally, we conducted a rigorous evaluation of the generator component, which 
typically consists of diffusion models in text-guided reconstruction 
methods. To ensure that a visual image reconstruction method has the potential 
to faithfully reproduce an individual's perceived visual experiences, it is 
crucial that the method can recover the original images with a high degree of 
perceptual similarity when the neural translation from brain activity to 
latent features is perfect. However, it has remained unclear whether recent text-guided reconstruction methods meet this fundamental requirement. To address 
this question, we performed a recovery check analysis by reconstructing images 
using the true latent features of target images. Instead of using latent 
features translated from brain activity, we directly input the latent features 
derived from the target images into the generator. 

\begin{figure}[!t]
  \centering
    \includegraphics[width=1.0\linewidth]{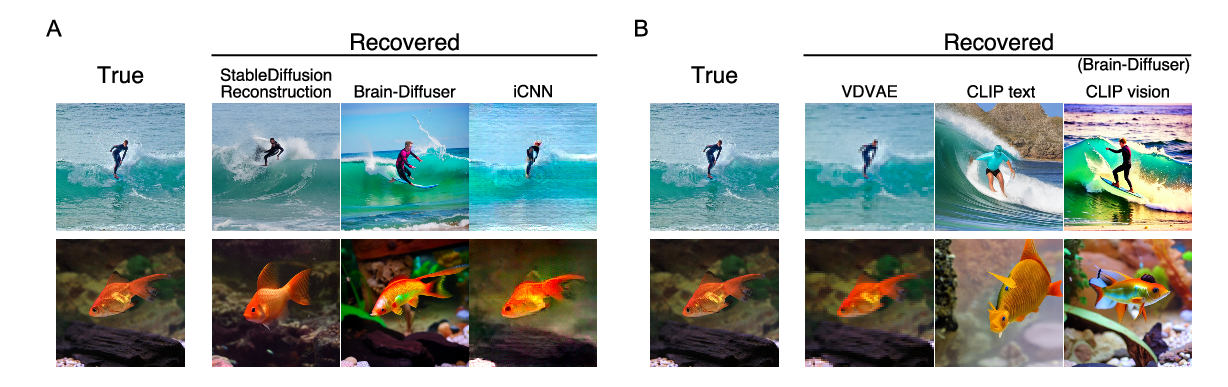}
  \caption{{\textbf {Recovery check.}} 
  (\textbf{A}) Reconstruction from the true latent features. The leftmost 
  column displays the original images. Subsequent columns show reconstruction 
  results from the StableDiffusionReconstruction, Brain Diffuser, and iCNN methods. 
  (\textbf{B}) Component-wise reconstruction of the Brain-Diffuser method. 
  The leftmost column shows the original images. The following columns present 
  reconstruction results using individual components of the Brain-Diffuser 
  method (VDVAE, CLIP text, and CLIP vision features).  Each row represents 
  a different test image. The reconstruction results indicate the upper bound 
  of reconstruction performance for each method and highlight potential 
  limitations in the latent feature representations or generative processes.
  }
  \label{Figure_07}
\end{figure}

Text-guided methods produced images semantically similar to targets but not perceptually similar (Fig.~\ref{Figure_07}A), while the iCNN method yielded results that 
closely resembled the actual target images. These findings suggest that text-guided reconstruction methods may prioritize semantic similarity over perceptual accuracy. In contrast, the iCNN method appears to have a superior ability to capture and replicate original visual content, indicating potential advantages in preserving fine-grained visual details.

To investigate the recovery performance further, we conducted a recovery 
check on each latent feature of the Brain-Diffuser method (Fig.~\ref{Figure_07}B). 
Interestingly, reconstructions from VDVAE features, which are used for 
generating initial images in the Brain-Diffuser, exhibited a high degree of 
similarity to the target images. However, the images generated by CLIP features 
through the diffusion models showed significant deviations from the original 
targets. These findings suggest that text-guided reconstruction methods may not 
be well-suited for visual image reconstruction tasks, as they fail to 
faithfully recover the original visual images. Instead, they tend to create 
images based on their \rev{semantic} features, such as CLIP features, which can lead 
to a phenomenon known as ``hallucination'' in the field of generative AIs. 
Hallucination refers to an output that appears plausible but is actually 
incorrect or misleading, raising concerns about the reliability and accuracy 
of the model \citep{rawte_survey_2023}. Text-guided reconstruction methods seem to prioritize 
generating semantically similar images rather than faithfully reconstructing the visual content perceived by the individual \rev{(see Fig.~\ref{Figure_A8} for the recovery with MindEye2’s generators)}.

The above findings may provide an explanation for why the text-guided 
reconstruction methods performed well only on the NSD dataset (Fig.~\ref{Figure_02}A). The results 
of the case study demonstrated that the text-guided reconstruction methods 
struggle to reconstruct (Fig.~\ref{Figure_02}B) or identify (Fig.~\ref{Figure_06}D) 
test samples that lie beyond the distribution of the training set. Such 
limitations suggest that these methods lack true generalization capabilities 
and are unable to accurately reconstruct novel visual experiences that differ 
significantly from the examples they were trained on. Moreover, even when the 
test samples belonged to the same distribution as the training set, the 
translators had difficulty correctly 
identifying those test samples (Fig.~\ref{Figure_05}). This observation indicates 
that the translators may not have learned a sufficiently robust and 
generalizable mapping between brain activity patterns and the corresponding 
latent features, further limiting their ability to faithfully reconstruct the 
perceived visual experiences. 

This case study also revealed that the NSD test stimuli are highly similar to the 
training set, with a significant overlap in their visual 
and semantic features (Fig.~\ref{Figure_04}). Given this similarity, the impressive reconstruction 
results achieved by the recent text-guided reconstruction methods on the NSD 
dataset should not be interpreted as evidence of zero-shot reconstruction 
capabilities. Instead, a more plausible interpretation is that these methods 
primarily function as a combination of ``classification'' and ``hallucination.''

In this context, the CLIP feature translator in the text-guided reconstruction 
methods functions primarily as a classifier, predicting categorical semantic information 
present in the training phase rather than capturing the fine-grained details of 
the visual experience. This limitation may explain why convincing reconstructions 
can be obtained even from random brain data through post-hoc selection (Fig.~\ref{Figure_03}BC). 
Due to the limited variety in the outputs generated by the reconstruction models, 
repeated trials and post-hoc selection can eventually find images that are 
semantically and visually similar to the target stimulus. The apparent 
plausibility and semantic consistency of these generated images can be 
attributed to the capabilities of the diffusion model, which learns to generate 
realistic-looking images based on semantic information. While these images 
may seem convincing at first glance, they do not accurately reflect the 
specific visual experience of the individual. This phenomenon of hallucination 
raises serious concerns about the reliability and validity of the text-guided 
reconstruction methods when evaluated on the NSD dataset. Although we particularly evaluated three reconstruction methods in this case study, it is important to recognize 
that any reconstruction methods evaluated only by NSD \citep{scotti_reconstructing_2023,Quan_2024_CVPR} 
can potentially have similar classification and hallucination problems \rev{due to the limited diversity of the NSD. Furthermore, recent highly realistic reconstruction methods leveraging generative AI models \citep{chen_seeing_2023,bai_dreamdiffusion_2024,benchetrit_brain_2024} should also be carefully validated to ensure their performance is not overly dependent on dataset biases (Fig.~\ref{Figure_A5}).}

\subsection{Formal analysis and simulation} 
Building on the issues identified in our case study, we examine these challenges in a more general context. We have identified several issues with these 
text-guided reconstruction methods and the dataset, including the cluster 
structure of CLIP latent features, the lack of diversity in the NSD, and the 
misspecification of the latent representation for image reconstruction. These issues 
resulted in the inability of diffusion models to faithfully recover the original 
images from their latent features. However, it is crucial to recognize that 
the findings of the case study are not merely specific to CLIP, NSD, or 
diffusion models. Instead, these issues likely reflect more fundamental 
problems that can arise in the development and evaluation of brain decoding 
and visual image reconstruction methods. Thus, in this section, we extend the 
problems identified in the case study into formal analyses and simulations in 
generalized settings, aiming to provide a more comprehensive understanding of 
the factors that contribute to the limitations of current reconstruction 
methods and explore strategies for mitigating these issues. 

\subsubsection{Output dimension collapse}
Multivariate linear regression models are widely used in constructing decoding and encoding models of the brain. These models are often regarded as capable of independent and compositional predictions for each target, as they create separate regression models for individual targets without sharing the weights between them. However, this expectation is not generally true. This is particularly 
evident when input variables are shared \citep{seeliger_generative_2018,ozcelik_reconstruction_2022,mozafari_reconstructing_2020, takagi_high-resolution_2023,ozcelik_natural_2023}. 
In this section, we demonstrate this assertion by examining the problem of predicting multiple targets 
using linear regression models with shared inputs.

Let us consider predicting a feature vector 
$\mathbf{y} \in \mathbb{R}^{D}$ from a brain activity pattern 
$\mathbf{x} \in \mathbb{R}^{D}$ using a linear regression model. For the 
training set, we consider brain activity matrix $X_\mathrm{tr}\in \mathbb{R}^{N \times D}$
and feature value matrix $Y_\mathrm{tr}\in \mathbb{R}^{N \times D}$, where $X_\mathrm{tr}$
consists of $N$ samples of $D$-dimensional brain activity vectors $\mathbf{x}$ and 
$Y_\mathrm{tr}$  consists of $N$ feature vectors $\mathbf{y}$. We then 
train a linear (ridge) regression model using this training data. Given a 
regularization parameter $\lambda$, the weight of the ridge regression model 
is analytically derived as $W = (X_\mathrm{tr}^{\top} X_\mathrm{tr} + \lambda I)^{-1} X_\mathrm{tr}^{\top} Y_\mathrm{tr}$
where $I$ is the $D \times D$ identity matrix. The predicted feature vector
$\hat{\mathbf{y}}_\mathrm{te}$ for the test brain activity data $\mathbf{x}_\mathrm{te}$ can be represented as:
\begin{align}
	\hat{\mathbf{y}}_\mathrm{te} &= W^{\top} \mathbf{x}_\mathrm{te} \\
	&= Y_\mathrm{tr}^{\top}X_\mathrm{tr} (X_\mathrm{tr}^{\top}X_\mathrm{tr} + \lambda I)^{-1}\mathbf{x}_\mathrm{te} \\
	&= Y_\mathrm{tr}^{\top} \mathbf{m} = \sum_i^N m_i \mathbf{y}_\mathrm{tr}^{(i)}, \label{eq:sum}
\end{align}
where ${\bf m} = X_{\rm tr} (X_{\rm tr}^{\top}X_\mathrm{ tr} + \lambda I) ^{-1}\mathbf{ x}_\mathrm{ te} \in \mathbb R^{N}$, $m_{i}$ is the $i$th element of $\mathbf {m}$, and $\mathbf{y}_\mathrm{ tr}^{(i)}$ 
is the $i$th training feature vector. This transformation indicates that 
the predicted value is always represented as a linear combination of the 
target features in the training set. This property is not limited to ridge 
regression but generally applies to ordinary ridgeless linear 
regression and related linear models.

Next, we consider a scenario where the diversity of the target features is 
small. This situation can arise when the feature space exhibits a clustered 
structure and the training data lacks sufficient diversity, as observed in the 
case study with the CLIP text features and the NSD dataset (Fig.~\ref{Figure_04}A). When the training 
features have limited diversity, the predicted values from brain 
activity, which are represented as linear combinations of these target features, 
also become constrained. Consequently, the prediction from brain data to 
target features effectively becomes a projection onto a low-dimensional subspace formed 
by the training data. 

\begin{figure}[!b]
  \centering
    \includegraphics[width=1.0\linewidth]{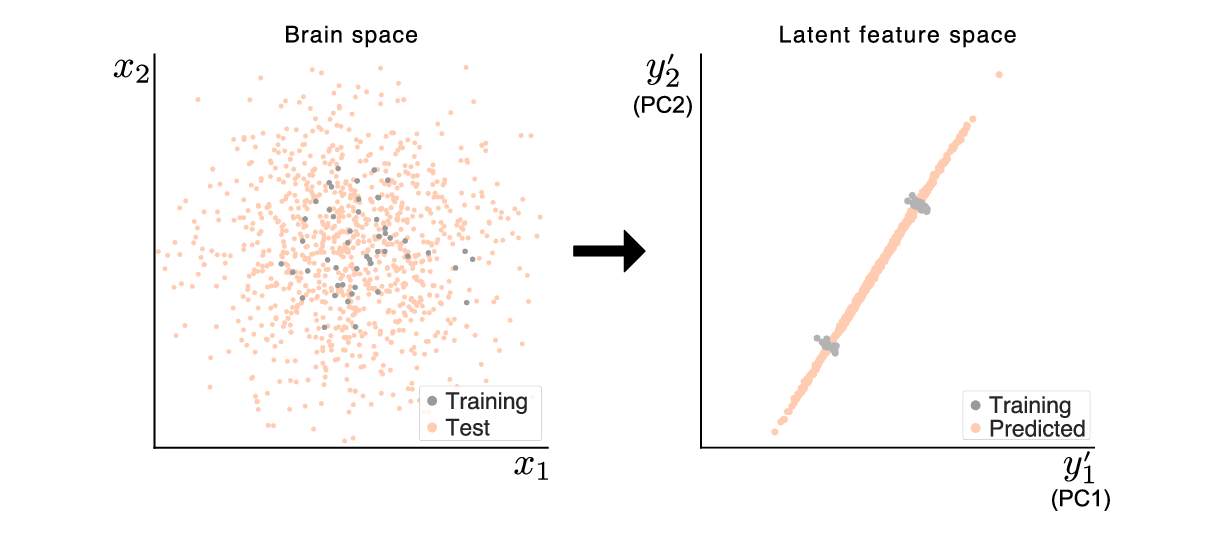}
  \caption{{\textbf {Demonstration of output dimension collapse in feature prediction.}} 
  The left panel shows the distribution of source brain activity data in the 
  first two dimensions of the original high-dimensional space. The right panel 
  displays the distribution of target latent features projected onto the first 
  two principal component (PC) dimensions. A linear ridge regression model is 
  trained using the training data (gray points in the left and right panels), 
  in which the output features in the training data are constrained to two 
  clusters. When presented with test data from the brain space (left panel), 
  the model generates predictions (orange points) in the latent feature space 
  (right panel). This visualization demonstrates how the predicted features are 
  constrained to the subspace defined by the training data, highlighting the 
  limitation in generalizing beyond the training data distribution.
  }
  \label{Figure_08}
\end{figure}

To illustrate this phenomenon, we simulated teacher-student learning, a framework where a ``teacher'' model generates data based on certain underlying rules, and a ``student'' model is trained to learn or approximate those rules by observing the generated data. We examined the distribution of predicted values from the student linear 
regression model trained on clustered features generated by the teacher model. We generated clustered features 
by sampling from a Gaussian mixture distribution in a high-dimensional space. 
The corresponding brain activity samples were generated from the latent feature samples 
multiplied by teacher weight and adding observation noise (see Methods ``Simulation with clustered data'').  Student weights were obtained by training a linear regression model to predict the clustered feature values from the corresponding brain data. \rev{We 
then projected the randomly generated brain samples into the latent feature space 
using the learned regression weights}. To visualize the 
high-dimensional predicted patterns effectively, we projected the predicted 
features onto the PCA spaces derived from the training 
features and showed the first two dimensions.

The simulation results clearly demonstrate the impact of clustered features on 
the predicted values (Fig.~\ref{Figure_08}). The trained linear regression model projects arbitrary 
brain data onto the subspace defined by the latent features in the training set, resulting in 
predicted values that are confined to the vicinity of the training clusters. 
This observation highlights the limitation of training linear regression models 
with clustered features as prediction targets: the trained model's predictions are inherently 
constrained by the diversity and structure of the training data.

We refer to this phenomenon as ``output dimension (or domain) collapse,'' where the model's predictions become confined to a limited subspace (or subdomain) of the output feature space. It has 
important implications for the generalization capability of linear regression 
models in the context of brain decoding and visual image reconstruction. When 
training data lack diversity and form distinct clusters in the feature 
space, the translator overly adapts to the subspace formed by the training 
data, regardless of the potential of the latent feature space. Consequently, 
the translator's outputs become confined to patterns similar to those in the training set, severely limiting the model's ability to predict 
novel or out-of-distribution samples. 

Output dimension collapse may explain why plausible reconstructions were obtained even from random brain data by merely generating images several times in the case study (Fig.~\ref{Figure_03}). The lack of semantic diversity in the NSD causes the translator to adapt only to the feature patterns of the training set, restricting its outputs to the subspace formed by the training data. As a result, convincing images could be found even from random data through questionable post-hoc selection. 

\rev{It should be noted that this phenomenon is not inherently limited to linear regression models; it can occur in various multivariate regression models, including multi-layer neural networks. In fact, when we replaced the nonlinear translator in the MindEye2 reconstruction with a linear translator, the reconstructions’ bias toward object categories in the training set was substantially reduced (Fig.~\ref{Figure_A9}). This observation suggests that nonlinear models are more susceptible to the collapse of the output domain due to their greater flexibility in fitting to training data.}

It is also important to recognize that the mathematical formulation in Eqs. 1--3 assumes that all input variables are shared across target variables. If each target is predicted from a distinct set of input variables through feature (voxel) selection, the predictions can become more independent, potentially mitigating output dimension collapse. This approach has been utilized in the field since its early days, with techniques such as sparse voxel selection and modular modeling \citep{miyawaki_visual_2008,Yamashita2008SparseEA, Fujiwara2013Modular, shen_deep_2019}. 

\subsubsection{Simulation with clustered features: What makes prediction compositional?}

The case study revealed that the NSD exhibits limited diversity 
(Fig.~\ref{Figure_04}) and poses difficulties for zero-shot prediction 
(Fig.~\ref{Figure_05} and Fig.~\ref{Figure_06}D). These observations suggest 
that the translator of CLIP features suffers from output dimension collapse 
due to the lack of semantic diversity in the NSD. To explore potential strategies 
for mitigating output dimension collapse and achieving flexible predictions, 
we conducted simulation analyses using clustered features to assess generalization 
performance beyond the training set. 

\begin{figure}[!p]
  \centering
    \includegraphics[width=1.0\linewidth]{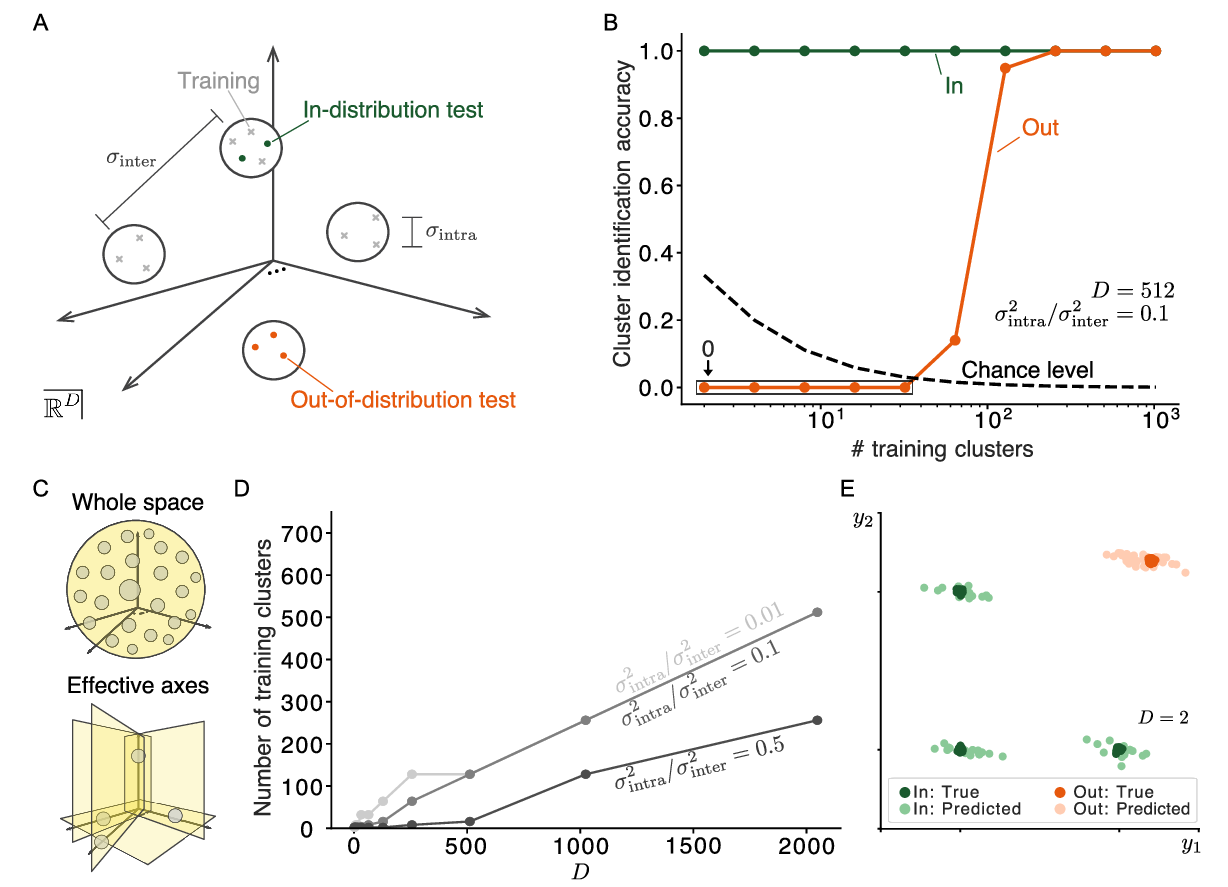}
  \caption{{\textbf {Simulation analysis of predicting cluster-structured features.}} 
   (\textbf{A}) Illustration of target latent features. The latent 
   features were generated from Gaussian mixture distributions, with $\sigma_{\mathrm {intra}}$ controlling within-cluster spread and $\sigma_{\mathrm {inter}}$ controlling inter-cluster scaling. In-distribution test samples are generated 
   from training clusters while Out-of-distribution (OOD) test 
   samples come from novel clusters. 
   (\textbf{B}) Cluster identification accuracy for various numbers of training clusters. The $x$-axis represents the 
   number of training clusters, and the $y$-axis shows the cluster 
   identification performance. The green and orange lines indicate results for in-distribution and OOD samples, respectively. The dashed curve indicates the chance level. (\textbf{C}) Scenarios for achieving generalizability with sufficient data diversity. The 
   upper illustration shows the training data covering the whole latent feature 
   space, requiring an exponential order relative to the feature dimension. 
   The lower one shows that the training data covers only the effective 
   axes of the latent feature space, leading to a linear order relative to the 
   feature dimension. (\textbf{D}) Sufficient number of clusters for generalization as a function of latent feature dimension. The $x$-axis represents 
   the dimension of the target latent features. The $y$-axis shows the number of 
   clusters achieving above $0.5$ cluster identification accuracy, with curves for different $\sigma_{\mathrm {intra}}^2/\sigma_{\mathrm {inter}}^2$ ratios. 
   (\textbf{E}) Example of successful prediction beyond the training distribution 
   in a 2D output feature space. In-distribution and OOD target features are depicted in dark green and dark orange, and the model's predictions of these features are depicted in light green and light orange, respectively. Despite the OOD target features (dark orange) not being included in the training clusters, they are accurately predicted (light orange), demonstrating the model's 
   generalization ability by combining learned feature dimensions.
  }
  \label{Figure_09}
\end{figure}

As in the previous section, our simulation involved teacher-student learning. We first generated 
feature data ${\mathbf y} \in \mathbb R^D$ then made the input brain data ${\mathbf x} \in \mathbb R^D$ by translating ${\mathbf y}$  with the teacher weight and added observation noise. To simulate a situation where the dataset has cluster structures and to control 
diversity effectively, the training feature vector $\mathbf{y}\in \mathbb R^D$
was generated from a $D$-dimensional Gaussian mixture (Fig.~\ref{Figure_09}A).

We trained a ridge regression model on large training data samples and obtained 
the student weight. To simulate the situation where the trained model encounters 
clusters that are not available at the training phase, we used two types of test 
samples, in-distribution and out-of-distribution test samples: in-distribution test samples 
were generated from one of the clusters used in the training set, whereas 
out-of-distribution (OOD) samples were generated from the novel cluster that is not included in the training set. 
For these two types of predicted features, we calculated the cluster identification accuracy 
(Fig.~\ref{Figure_06}C) by using $C + 1$ cluster centers: $C$ centers from the training 
set and one cluster center of the OOD test set.

We first examined zero-shot prediction performance for different numbers of training clusters (Fig.~\ref{Figure_09}B). We fixed the feature dimension $D$ and the cluster variance ratio constant. While the cluster identification performances of in-distribution test samples were perfect, the performances of OOD test samples showed different patterns depending on the number of clusters in 
the training data. When the number of training clusters was small, the cluster 
identification accuracy was $0$\%. This result indicates that the behavior of the 
translator became more similar to that of a classifier, making it difficult to generalize beyond the training set, as observed in the NSD cases (see Fig.~\ref{Figure_06}D). 
On the other hand, as the number of clusters in the training data increased, 
it became possible to identify the novel clusters, achieving the same 
performance as in-distribution test samples. This observation indicates the 
importance of the diversity of the training dataset. We also emphasize that large numbers of training samples do not necessarily address the problem. 
All of these results were obtained with a sufficiently large amount of training data, and the number of training clusters was varied while keeping the amount 
of training data fixed. Also, we observed qualitatively similar results in 
increasing the data diversity by controlling the cluster variance ratio while 
keeping the number of dimensions and training clusters constant (Fig.~\ref{Figure_A10}A).

Next, we investigate how diverse the training data needs to be to ensure 
sufficient generalization. There are two possible scenarios for diversifying 
training samples: either by densely sampling the entire target feature space so that there 
are no remaining gaps  (Fig.~\ref{Figure_09}C; top) or by uniformly sampling 
to the extent that it covers the entire dimension of the target feature space 
(Fig.~\ref{Figure_09}C; bottom). The former scenario requires an exponentially 
larger number of samples/clusters relative to the dimension, whereas the latter
only requires up to a linear order. We sought to reveal which of these two 
scenarios was more likely to be true by varying the dimensions of the feature 
space and identifying the number of clusters required for generalization 
(Fig.~\ref{Figure_A10}B for identification accuracy in each condition). Here, we defined the number of clusters required for 
generalization as the point at which the identification accuracy of OOD samples 
exceeds $50$\%. The relationship between the dimension of the feature space 
and the number of clusters necessary for generalization appears to be 
linear (Fig.~\ref{Figure_09}D). This finding suggests that achieving 
generalization does not necessarily require an exponentially large diversity 
that fills the entire feature space. Instead, it suffices to have a number of 
clusters that cover the adequate dimensions within the target feature space. 
Although obtaining a large amount of brain data is hard work, it is important 
for a dataset to contain sufficiently diverse stimuli covering the effective 
dimensions of the target feature space to achieve zero-shot prediction. 

We also confirmed this phenomenon with a simple and transparent example 
($D = 2$, Fig.~\ref{Figure_09}E). The training data covers sufficient axes in 
the target feature space, enabling the prediction of locations not present in the 
training set. Based on this low-dimensional intuition, we argue that successful 
zero-shot prediction requires training data to leads representations that can serve as a basis for spanning the target feature space. Leveraging such 
bases effectively enables the model to predict novel samples by predicting 
each basis and combining them. This compositional representation, spanning 
the target feature space, is crucial for zero-shot prediction 
\citep{schug_discovering_2024} and reconstructing arbitrary visual images 
from limited brain data. 

\revrev{While our simulations varied the number and spread of clusters to examine the role of training diversity, we assumed that cluster centers were uniformly distributed across the latent feature space. However, this assumption may not accurately reflect the characteristics of actual datasets such as NSD, where cluster centers themselves can potentially be biased in the entire visual space (\textit{i.e.}, focusing on natural scenes in MSCOCO). Even with a large number of clusters,  generalization to unseen images remains challenging if they occupy only a limited region of the visual space. These considerations emphasize that true diversity requires careful control over both the number and the spatial distribution of training clusters or images.
}

\subsubsection{Caveat with evaluation by pairwise identification}

Pairwise identification has been a standard metric for evaluating latent 
feature decoding \citep{horikawa_generic_2017} or reconstruction performance \citep{beliy_voxels_2019, shen_end_end_2019,
mozafari_reconstructing_2020,qiao_biggan-based_2020,ren_reconstructing_2021,gaziv_self-supervised_2022,
takagi_high-resolution_2023,ozcelik_natural_2023,scotti_reconstructing_2023,
denk_brain2music_2023,koide-majima_mental_2024}. However, our analysis revealed 
that even with difficulties in accurately identifying specific semantic clusters the 
test samples belong to, the pairwise identification performance still surpassed 
its chance level (Fig.~\ref{Figure_06}F). \rev{This result highlights a fundamental limitation of pairwise identification: its susceptibility to overestimation when the dataset or target features contain a strong high-level categorical structure.} Here we critically examine this metric 
and demonstrate that significant results can be easily obtained when the 
target or predicted features exhibit certain structures. 

Pairwise identification is calculated as the accuracy with which the predicted 
features (either the output of a translator or features extracted from the output of a generator) can correctly identify the corresponding true 
ones, in pairs consisting of a true sample and one of the remaining samples 
in the test set. We refer to the latter remaining sample as a candidate sample in the following. If the candidate sample belongs to the same category as 
the true sample, the identification is expected to be difficult. 
Conversely, if the candidate sample belongs to a different category from the 
true sample, identification becomes easier. \rev{This characteristic makes the metric highly dependent on categorical distinctions rather than the actual quality of feature prediction or reconstruction.}

\begin{figure}[!t]
  \centering
    \includegraphics[width=1.0\linewidth]{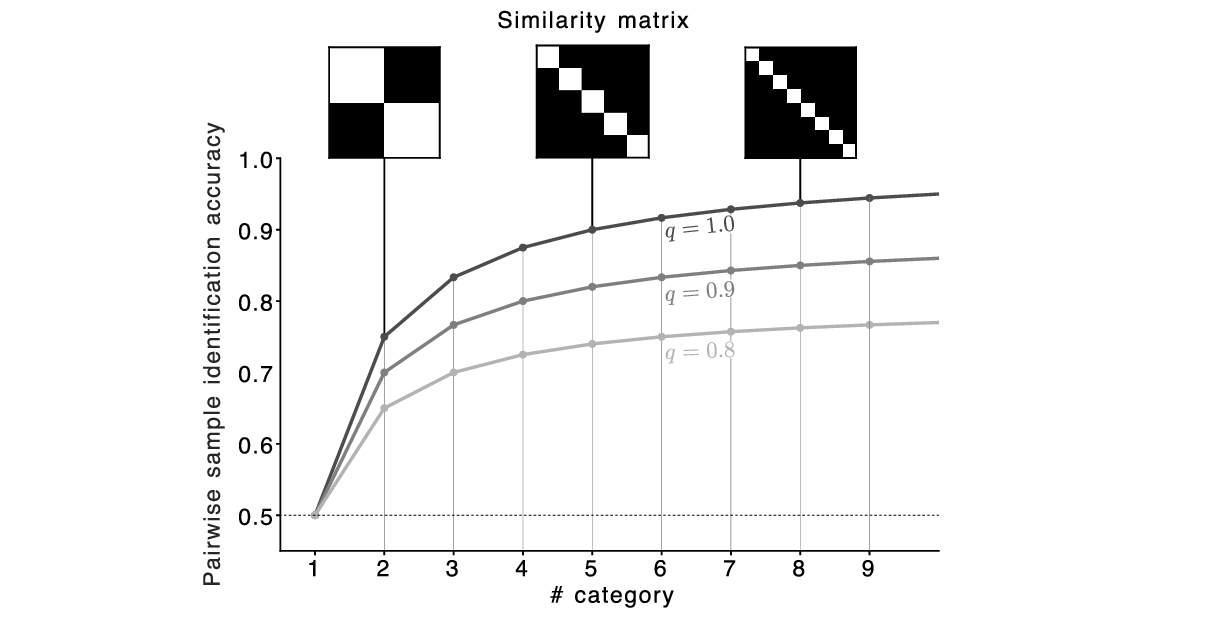}
  \caption{{\textbf {Expected pairwise sample identification performance in categorically structured data.}} 
  The $x$-axis represents the number of categories in the test set. The $y$-axis 
  represents the pairwise sample identification accuracy. Different lines 
  represent various levels of accuracy ($q$) in distinguishing samples from 
  different categories. The chance level of $0.5$ is represented by the bottom 
  line.  Above the graph, hypothetical similarity matrices are shown to 
  illustrate the categorical structures of the samples, where the color 
  (white/black) indicates similarity between samples. The block diagonal 
  structure reflects the categorical nature of the data. Samples within the 
  same category are assumed to be indistinguishable, resulting in a pairwise 
  identification accuracy of $0.5$ (chance level). Samples from different 
  categories can be distinguished with a pairwise identification accuracy of $q$, 
  where q varies between $0.5$ and $1$.
  }
  \label{Figure_10}
\end{figure}

Here, we assume the test set comprises $k$ categories with test samples equally distributed across each category for simplicity. We model the situation 
mentioned above by setting the expected identification accuracy at the chance level 
(\textit{i.e.}, $0.5$) when a candidate sample belongs to the same category as the true one. 
Conversely, when the candidate sample belongs to a different category from the true sample, we set the expected identification accuracy to a parameter $q$, ranging from $0.5$ to $1$. This parameter $q$ reflects the ease of identification across categories. Assuming the number of 
test samples is sufficiently large, the pairwise identification accuracy $\mathrm{Acc}$  becomes
\begin{equation}
	\mathrm{Acc} = \frac{1}{k} \cdot 0.5 + \left(1-\frac{1}{k}\right) \cdot q
\end{equation}
(see Methods ``Expected identification accuracy 
in imprecise reconstructions'' for the derivation).

Fig.~\ref{Figure_10} illustrates the relationship between pairwise identification accuracy and the number of categories in the test set through line plots of expected values. Inset figures depict the underlying similarity structure of the test set. Notably, even when identification within categories fails completely and succeeds only between two categories, pairwise identification accuracy can still reach a high value of up to $75$\%. \rev{This highlights a major limitation: above-chance performance may simply reflect an ability to differentiate broad categories rather than accurately reconstruct crucial visual details. Indeed, the same pattern emerges even under a hold-out split when the samples in the held-out cluster can still be categorized broadly. Consequently, performance evaluation only relying on this metric can lead to misleading conclusions about the model’s actual capabilities and generalizability.}

\subsubsection{Preserved image information across hierarchical DNN layers}

The reconstruction of arbitrary visual images requires compositional latent features that can be effectively mapped into the image space. As our case study has suggested, hierarchical DNN features from VGG19 have been found to be suitable for zero-shot prediction or reconstruction tasks due to their compositional representations. At the same time, however, the extent to which these features truly map to the image space remains unclear. Indeed, a common narrative suggests that the hierarchical processing discards pixel-level information through progressively expanding receptive fields. Yet, this view is not entirely accurate.

\begin{figure}[!p]
  \centering
    \includegraphics[width=1.0\linewidth]{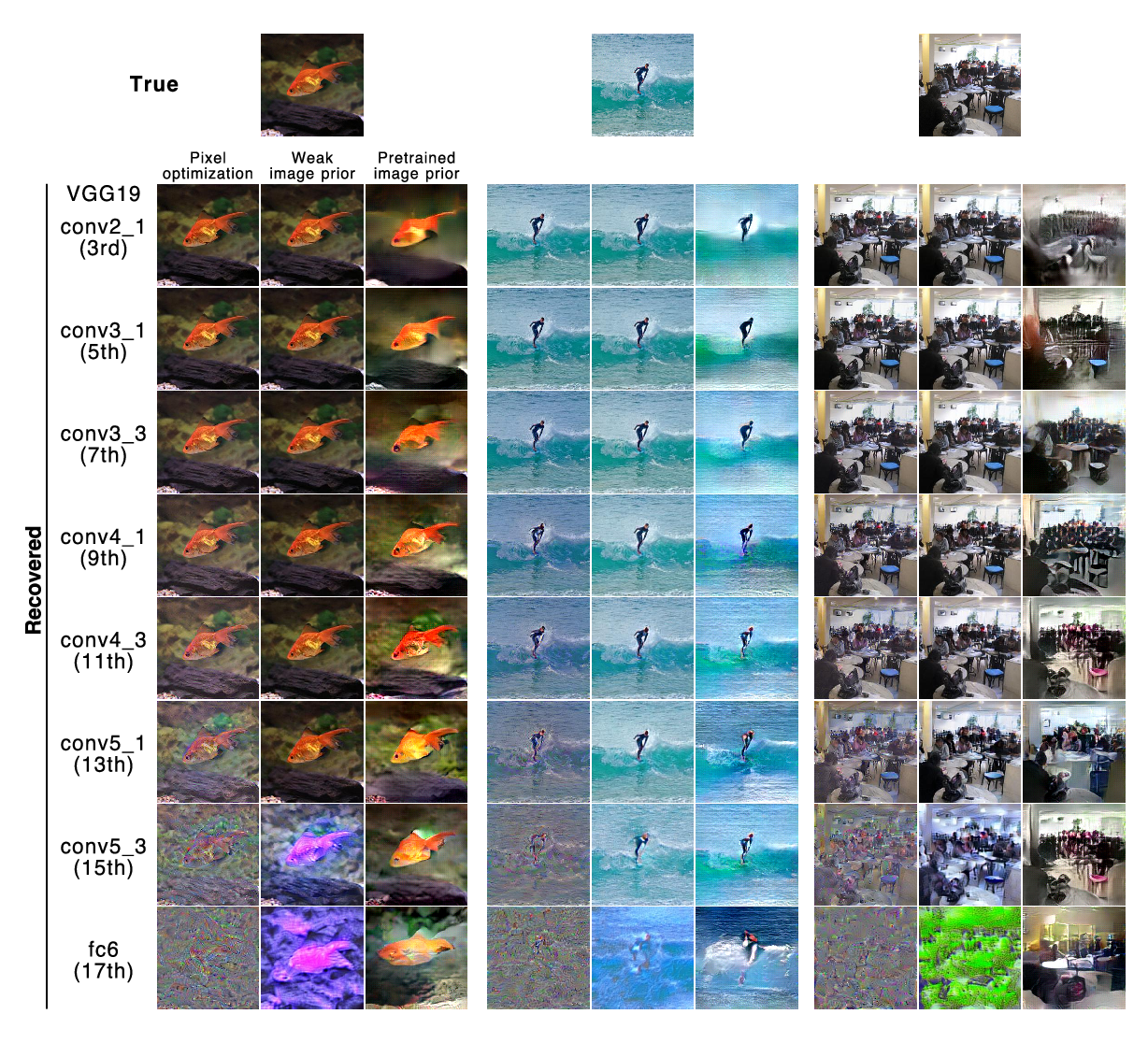}
  \caption{{\textbf {Image recovery check for hierarchical latent features.}} 
  The images show the ability to recover visual inputs from different layers of 
  the VGG19 network. For each of the three sample images, three different 
  optimization methods were applied. The left column shows results from pixel 
  optimization, which directly optimizes pixel values to minimize the loss 
  between the generated image features and the target image features \citep{mahendran_understanding_2015}. 
  The middle column displays optimization with a weak image prior, 
  simultaneously optimizing both the weight parameters of the image prior models and 
  their latent features \citep{ulyanov_deep_2017}. The right column presents 
  optimization with a pre-trained image prior, which optimizes the latent 
  features of a parameter-fixed image generator model to minimize the loss 
  between the output image features and the target image features \citep{dosovitskiy_generating_2016}. 
  Rows correspond to different layers of the VGG19 network, progressing from 
  earlier layers at the top to later layers at the bottom. This progression 
  illustrates how image information is preserved or lost at different stages of 
  the network. Reasonable image recovery is possible even from relatively 
  high-level layers of the VGG19 network, challenging the notion that higher 
  layers in deep neural networks discard image-level visual information.
  }
  \label{Figure_11}
\end{figure}

For example, the latent features of auto-encoder models \citep{hinton_reducing_2006,kingma_auto-encoding_2014,oord_neural_2018}
can represent images in a low-dimensional space while preserving their 
reversibility, which is reasonable considering that the model's output is trained to 
match the input. Furthermore, \cite{mahendran_understanding_2015} 
showed that input images can be recovered with reasonable accuracy even 
from relatively high-level layers of a DNN designed for an object recognition 
task. It has also been argued in the neuroscience field that large receptive 
field sizes do not necessarily impair neural coding capacity as long as the 
number and density of units remain constant \citep{zhang_neuronal_1999,majima_position_2017}. 
These results challenge the notion that higher-level layers in DNNs discard all 
pixel-level information. 

To further illustrate 
this point, we performed a recovery check on each intermediate layer of the 
VGG19 network used in the iCNN methods (Fig.~\ref{Figure_11}; see also Fig.~\ref{Figure_07}). 
Given the DNN features of a target image, we optimized input pixel values to 
make the image's latent features similar to the targets (see Methods ``Recovery check of a single layer by iCNN''). We observed that input 
images can be recovered with reasonable accuracy from relatively high-level 
layers (around the 11th layer out of the total 19th layers). Furthermore, by 
introducing image generator networks to add constraints on image statistics, 
reasonable recovery can be achieved from even higher layers. By utilizing a
weak image prior \citep{ulyanov_deep_2017}, which contains only information about 
the structure of images without any prior information on natural images, input 
images can be recovered from the 13th layer. When using an image generator 
that has learned natural image information \citep{shen_deep_2019,dosovitskiy_generating_2016}, 
input images can be recovered even from the 15th layer.

These observations suggest that, even when feature representations shift from lower to higher levels through hierarchical processing, pixel-level information is not largely discarded; rather, much of the input information is preserved across almost the entire level. This perspective highlights the potential for utilizing intermediate DNN representations in the visual image reconstruction study as the generator should recover the original stimulus from the true latent features (see also “Failed recovery of a stimulus feature from its latent features” in the case study section). With this insight in mind, exploring which representations have compositional representation and are more predictable from brain activity will be a critical step in advancing visual image reconstruction.

Conversely, utilizing high-level image features, such as the output of DNNs, or features from other modalities, such as text annotations, is not a rational choice for visual image reconstruction tasks. These latent features make it challenging to recover the corresponding input image (Fig.~\ref{Figure_07}B) and are insufficient as surrogate representations of perceived visual images. Although recent text-to-image models and predicted text features can easily generate images, the outputs should not be interpreted as reconstruction results. Instead, it is more appropriate to view them as visualizations of decoded semantic information. While such visualizations are valuable for illustrating purposes, it is crucial to recognize the significant distinction between semantic visualization and reconstruction.

\subsubsection{How are we fooled by hallucinations of generative AIs?}

Generative AIs have recently made remarkable progress, with models now capable 
of producing high-resolution and realistic images from text input \citep{ramesh_zero-shot_2021} 
or generating text of a quality indistinguishable from human-written content \citep{brown_language_2020}. 
However, due to the complex internal structure of these models and the vast amounts of data 
they are trained on, we are often fooled by the outputs of generative AIs.
For instance, when searching for an unfamiliar topic using a large language model 
(LLM) in our daily lives, we may not realize that the model is creating false 
concepts. This is likely a result of generative AIs being trained on a large 
amount of data and producing highly coherent and contextual responses. We may 
also believe these models are unbiased and can represent all possible 
data points, even though they inherently contain biases from their training 
data and developers \citep{messeri_artificial_2024}. As we have observed, 
the generative AI-based reconstruction methods exhibit realistic appearances 
but poor generalizability (Fig.~\ref{Figure_02}). Could similar issues 
occur in visual image reconstruction studies as well?

The goal of visual image reconstruction is to generate images from brain 
activity that precisely mirror visual experience. However, there appears to 
be a prevalent focus among the general public, reviewers, and even researchers 
on achieving as realistic outputs as possible, rather than 
emphasizing the accuracy of these reconstructions. This shift in focus raises 
questions about the extent to which these realistic reconstructions truly 
represent the actual visual experiences.

Traditionally, we have held two beliefs: (1) generating realistic images from 
brain activity is challenging, and (2) if the reconstruction pipeline effectively 
captures the brain's representation under natural image perception, the model's 
output should also appear realistic. Based on these beliefs, realistic reconstruction is often considered an indication of accurately reflecting 
the actual visual experience.

\begin{figure}[!t]
  \centering
    \includegraphics[width=1.0\linewidth]{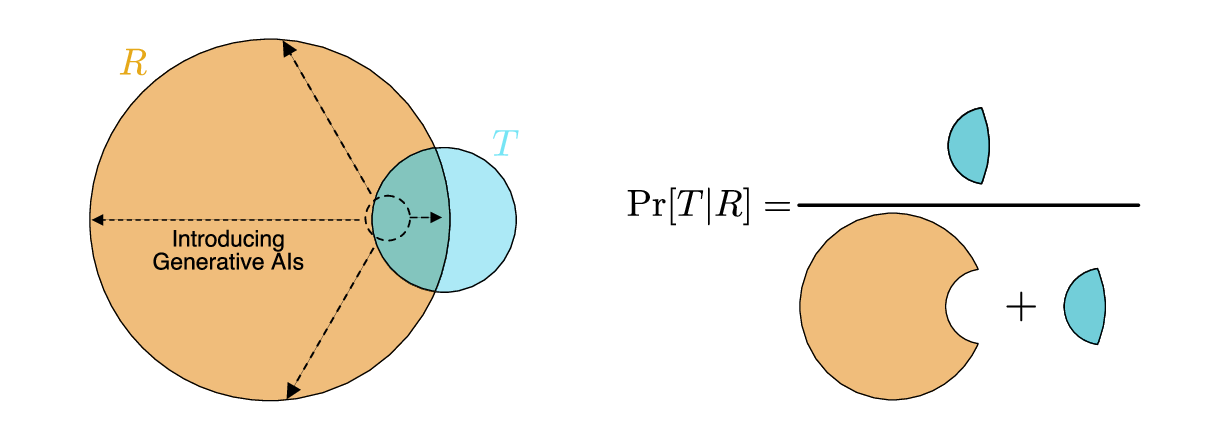}
  \caption{{\textbf {The illustration of how realistic appearances can be misleading in visual reconstruction.}} 
  On the left, Venn diagrams depict the relationship between two events: $R$, 
  where the output has realistic appearances, and $T$, where the output 
  truthfully reflects the visual experience. $T$ and a small dashed circle of $R$ 
  depict a common pre-generative AI heuristic: the assumption that realistic 
  appearance implies truthful reconstruction. Dashed arrows illustrate the 
  effect of generative AI models (\textit{e.g.}, diffusion models) on this relationship. 
  These models can potentially increase the overlap between $T$ and $R$, but 
  may also expand areas of $R$ unrelated to $T$. On the right, a probabilistic 
  interpretation of the relationship is presented. The equation represents 
  the probability that the generator's output truthfully reflects visual 
  experience ($T$), given that the generator's output has a realistic 
  appearance ($R$). This conditional probability is derived from the areas in 
  the Venn diagram. The expansion of $R$ without a proportional increase in $T$ 
  can lead to a decrease in $\Pr[T \mid R]$, emphasizing the need for careful 
  evaluation of reconstruction results beyond just assessing their realistic 
  appearance.
  }
  \label{Figure_12}
\end{figure} 

To formalize this heuristic reasoning, let us first define the two events $R$ and $T$, 
where $R$ represents the event that the output of reconstruction models has 
a realistic appearance, and $T$ represents the event that the model's output 
truthfully reflects the visual image. The first belief, concern about the difficulty of generating realistic images from brain activity, is expressed 
as $\Pr[R] \ll \Pr[\bar{R}] \approx 1$ where $\bar{R}$ represents the 
complementary event of $R$. The second belief, that reconstruction achieves 
a realistic appearance if the pipeline effectively captures the brain's 
representation under natural image perception, is expressed as 
$\Pr[R \mid T] \approx 1$. This conditional probability implies that the likelihood of the model’s output being realistic is high, given that the model truthfully captures the subject’s visual experience. The heuristic reasoning that realistic reconstructions indicate an accurate reflection of the actual visual experience can thus be represented 
as ``$\Pr[T \mid R] \text{ is high}$.'' This heuristic is, in fact, reasonable as 
it can be derived from Bayes' rule, assuming the above two beliefs hold true, 
and that $\Pr[T]$ is not extremely low.

However, recent developments in generative AIs, such as diffusion models, 
have made it easy to produce convincing, realistic outputs, subverting the 
first assumption, \textit{i.e.}, $\Pr[R] \gg \Pr[\bar{R}]$. Consequently, it becomes 
invalid to infer that the visual images are accurately reflected in the 
generator outputs solely because they appear realistic. Rather, as shown in 
Fig.~\ref{Figure_12}, the probability $\Pr[T | R]$ may become smaller as 
the generative AIs produce more convincing outputs (see also Fig.~\ref{Figure_07}). 
This perspective emphasizes the need for careful evaluation of reconstruction 
performance, considering the possibility of hallucinations by generators. 
While pursuing realistic reconstructions to improve reconstruction fidelity 
is undoubtedly important, it would be counterproductive to obsess over 
naturalistic appearance to the point of neglecting the original goal of 
reconstructing perceived visual images. 

\section{Discussion}

In this study, we critically examined generative AI-based visual 
image reconstruction methods to assess their true capabilities and limitations. 
Our primary goals were to (1) investigate the performance of these methods 
on different datasets, (2) identify potential issues and pitfalls in their 
methodology and evaluation, and (3) provide insights and recommendations for 
future research in this field. We conducted a case study focusing on text-guided reconstruction methods and their validation on the Natural Scene 
Dataset (NSD). Our findings revealed several concerns, including the failure 
to replicate the reconstruction performance on a different dataset, the use of 
problematic post-hoc image selection procedures, the lack of diversity and 
a limited number of clusters in the NSD stimulus set, the failure of zero-shot 
prediction by the translator component, and the inability to recover 
original stimuli by the generator component accurately. Formal analysis and simulations 
further demonstrated the phenomenon of output dimension collapse, the importance 
of compositional representations for achieving zero-shot prediction, and the 
potential pitfalls of relying solely on identification metrics to evaluate 
reconstruction performance. Moreover, we highlighted that a realistic appearance 
does not necessarily imply an accurate reflection of the perceived visual images. 
Based on these findings, we argue that the reconstructions from the recent 
text-guided reconstruction methods are, in large part, the result of a 
combination of classification and hallucination. Our study emphasizes the need 
for more rigorous evaluation and careful interpretation of results in visual 
image reconstruction research, particularly when using generative AI-based 
methods. 

\rev{While our study highlights the limitations of text-guided diffusion models for visual image reconstruction, it is important to acknowledge that these methods offer promising directions for brain decoding research. For instance, they can generate ``ROI optimal stimuli,'' which create images that activate a certain ROI activities maximally while not activating other ROIs activities, through the learned mapping between brain activities and latent features \citep{ozcelik_reconstruction_2022, ozcelik_natural_2023}. Although current studies focus on well-known functional ROIs (\textit{e.g.}, face, word, place, and body regions), this approach can extend to less understood brain regions. By generating optimal stimuli and identifying robust visual patterns, we can formulate new hypotheses about functional representations in previously uncharacterized brain areas. This data-driven approach complements traditional hypothesis-driven methods, potentially uncovering novel functional regions overlooked by conventional analyses.}

\rev{Moreover, while we emphasized the importance 
of zero-shot prediction, it is crucial to recognize that most brain 
decoding studies focus on classification tasks. Although these tasks are not zero-shot, 
they have nonetheless yielded valuable insights into neural representations \citep{haxby_distributed_2001,kamitani_decoding_2005}. In that sense, the text-guided or diffusion-based methods can also be utilized as tools for the visualization of decoded semantic contents (\textit{e.g.}, the supplementary movies in \citealp{horikawa_neural_2013}). Such visualization can be highly useful for visually conveying decoded information, even if it does not constitute zero-shot prediction.}

\rev{Additionally, it is worth noting that the individual components of text-guided reconstruction methods have already been utilized in various brain decoding applications. For instance, text latent features from deep neural networks and large language models (LLMs) have shown promise in analyzing semantic information from brain activity \citep{tang_semantic_2023,caucheteux_evidence_2023,Zhou2024CLIPMUSEDCM}. Furthermore, diffusion models can extend beyond text-to-image generation as MindEye2 generates images from visual latent spaces. Notably, \cite{cheng_reconstructing_2023} successfully reconstructed subjective experiences using a diffusion model within a carefully designed experiment. Leveraging these components and exploring their potential synergies, researchers can advance brain decoding and visual image reconstruction while addressing the challenges highlighted in our study.}

The recent trend of collecting and sharing large-scale visual neural datasets, 
such as those by \cite{hebart_things-data_2023} and \cite{xu_alljoined_2024}, is a welcome development in the field of 
neuroscience. These datasets provide valuable resources for researchers to 
investigate brain function and advance our understanding of visual processing. 
The NSD is a particularly notable example, as it was created with the goal of 
extensively sampling brain responses to a wide range of natural visual stimuli \citep{allen_massive_2021,naselaris_extensive_2021}. 
The NSD has been widely utilized in various studies \citep{prince_improving_2022,gifford_algonauts_2023, conwell_large-scale_2024}, 
demonstrating its value to the research community. While our results suggest that 
the semantic and visual diversity of the NSD stimuli may not be as high as 
initially thought, and there is substantial overlap between the training and 
test sets provided by the NSD authors, this does not diminish the overall 
importance and usefulness of the dataset. However, to fully leverage the NSD and 
other publicly available large-scale datasets for developing generalizable and 
zero-shot prediction models, it is crucial to consider the data split 
between training and test sets carefully. \revrev{While many large-scale datasets provide designated training and test splits, these splits are often not optimally designed to evaluate zero-shot prediction performance. When aiming for generalizable predictions beyond training examples as in visual image reconstruction, researchers should carefully verify whether significantly similar stimuli are included in both the training and test sets (Fig. \ref{Figure_A5}). If significant overlap exists, redesigning the training-test split becomes necessary to ensure the test set contains stimuli substantially dissimilar from those in the training set, thereby enabling genuine evaluation of generalization capabilities (Fig. \ref{Figure_A7}).} Moreover, recent 
advancements in functional alignment and inter-site neural code conversion 
methods \citep{haxby_common_2011,yamada_inter-subject_2015,wang_inter-individual_2024} hold 
promise for combining datasets from different sources, enabling truly 
larger-scale data analysis in neuroscience. These techniques allow researchers 
to align brain activity patterns across individuals and measurement sites even 
when stimuli are not shared across datasets. By leveraging these methods, 
researchers can pool data from various sources, increasing the sample size and 
diversity of the combined dataset, mitigating the limitations of individual 
datasets, and enhancing the development of generalizable and zero-shot 
prediction models.

Investigating neural responses to natural stimuli is a highly valuable approach 
to understanding brain function and representation \citep{nastase_keep_2020,hasson_direct_2020}. 
As our brains have developed while being exposed to natural scenes, it is 
crucial to use natural stimuli, especially in model training. However, we should 
not forget that we are also capable of perceiving non-natural stimuli like 
artificial images. We would like to emphasize that there are potential 
pitfalls when relying too heavily on evaluations based solely on natural stimuli. 
With the increasing scale of neural data and the growing complexity of analysis 
pipelines, there is a risk that the learned mappings may produce unexpected 
shortcuts, just as we have demonstrated that the text-guided reconstruction methods exploited the semantic and visual overlap between 
training and test sets. In the field of comparative and developmental 
psychology, researchers often prioritize using not natural but simple 
stimuli for better experimental controls and more precise inferences about 
infants' cognitive abilities \citep{kominsky_simplicity_2022, frank_baby_2023}. Drawing inspiration from this approach, we argue that the evaluation of visual image reconstruction should not be limited to complex natural stimuli alone. While natural stimuli are essential for ensuring ecological validity and understanding how the brain processes real-world information, it is equally important to assess the performance in controllable and transparent manners.

We have observed that an inappropriate split between training and test stimulus 
sets can lead to spurious reconstruction, invalidating zero-shot predictions. 
It is crucial to recognize that these problems can arise in various research 
contexts. Several studies have used data where the stimuli shared the category 
information between the training and test sets, potentially compromising the 
validity of their results. \cite{kavasidis_brain2image_2017} 
collected a dataset of EEG signals recorded during natural image perception. 
Their visual stimuli consisted of $2,000$ images selected from $40$ object 
categories in ImageNet ($50$ images per category) and the test set contained 
the same categories that are included in the training set 
(see also \cite{li_training_2018} and \cite{xu_alljoined_2024} for other issues with 
the data set). \cite{denk_brain2music_2023} attempted to develop 
music reconstruction methods from fMRI activity patterns. Their music stimuli 
consisted of $540$ music pieces selected from $10$ music genres, and the test 
set contained the same genres as the training set \citep{nakai_correspondence_2021}. 
\cite{orima_decoding_2024} attempted to reconstruct perceived 
texture images from EEG signals. Their texture stimuli consisted of $191$ image 
patches extracted from $21$ natural textures, and they performed a 
reconstruction analysis in a leave-one-out manner. It should be carefully 
examined whether these studies may suffer from output dimension collapse, merely decoding the broad category-level information observed in the training set.

Experimental design for training--test stimulus setups requires 
careful consideration. \cite{dado_brain2gan_2024} conducted an image 
reconstruction analysis from the multi-unit activity of a macaque using images 
generated from latent features of generative models. Importantly, all of the 
test stimuli were generated from the averaged latent features of the categories 
used in their training phase, \rev{suggesting the test stimuli are highly biased to the training set (see also Fig.~\ref{Figure_A5})}. While the authors addressed potential biases by 
redesigning the training and test split, researchers should exercise prudence when utilizing the dataset. Our inspection of the movie stimuli from \cite{nishimoto_reconstructing_2011} revealed that many frames 
in the test movie stimuli were nearly identical to those in the training set 
(Fig.~\ref{Figure_A11}). This similarity likely results from temporally adjacent 
video frames being split between the training and test stimuli. While our 
preliminary analysis of their latent features (motion energy features) did not 
show unusual clustering (note also that the study presents a way of retrieving 
movie instances via a brain encoding model, rather than reconstruction in the 
current sense), caution is required when using the stimulus set to extract 
other types of features \citep{huth_continuous_2012,huth_decoding_2016}. 
Claims of predicting arbitrary instances or achieving zero-shot prediction 
warrant thorough scrutiny. 

There is a widespread practice of using test data for fine-tuning, which can also be a questionable procedure. Some studies have proposed methods that involve fine-tuning models using the entire test brain data, but not the test stimuli, following initial training with the training stimuli and brain data \citep{beliy_voxels_2019,chen_seeing_2023}. While the absence of test labels (stimuli) during fine-tuning may help avoid obvious overfitting, these approaches still treat test brain data as previously observed information, rather than as a proxy for novel data potentially encountered in real-world situations. Consequently, this practice violates test data independence, making it difficult to evaluate the model's actual generalization ability. Although such procedures may effectively improve performance on existing benchmarks or competitions, it is crucial to recognize that incorporating test data information during any stage of training can undermine the validity of neuroscientific claims and limit the real-world applicability of the methods.

The issue of double dipping, which refers to the use of the same dataset for both data/variable selection and selective analysis (inference and prediction), has been widely recognized in neuroscience \citep{kriegeskorte_circular_2009,button_double-dipping_2019}. In classification tasks, while selecting input variables using test data is problematic, using the same output labels (target variables) for training and test sets is not inherently flawed, given that the nature of classification assumes consistent categories across datasets. However, the challenges we address in this study, although related to double dipping, present distinct concerns. In evaluating zero-shot prediction performance, the mere similarity of test labels (features) can lead to overestimating model performance. It is crucial to distinguish between conventional double dipping and the current issues we identify, such as output dimension collapse in zero-shot prediction scenarios. These emerging challenges necessitate careful consideration not only of input data independence but also of structural similarities between training and test sets in the output space.

Additionally, the independence of test stimuli from the model training processes requires careful examination, particularly when using pre-trained deep neural network (DNN) models and foundation models like CLIP or diffusion models. These models are typically trained on vast amounts of data available on the internet \citep{radford_learning_2021,brown_language_2020, rombach_high-resolution_2022}, which likely includes public datasets such as MS-COCO \citep{lin_microsoft_2014} that are used in the NSD. This overlap raises potential concerns about the true independence of test stimuli, as we cannot rule out the possibility that these pre-trained models have acquired representations specifically tailored to the stimuli used for the model training. To address these issues and ensure a more rigorous evaluation of model generalization capabilities, researchers may consider using test stimuli that are not publicly available on the internet. This could include self-created stimuli or carefully curated datasets that have not been used in the training of widely used AI models \citep{shen_deep_2019,cheng_reconstructing_2023}. Such an approach would provide a more stringent test of a model's ability to generalize to truly novel inputs.

\rev{Although we highlighted how stimulus overlap and limited diversity can lead to an overestimation of visual image reconstruction performance, such considerations do not apply to all reconstruction tasks. The dimensionality of the target space varies depending on the domain we seek to reconstruct; for instance, movement reconstruction often involves low-dimensional outputs (\textit{e.g.}, movement direction or velocity), where a small number of brain activity samples may suffice. In such cases, robust reconstruction is achievable by memorizing the brain-target pairs that cover the output space and classifying the test brain data into trained targets or interpolating between them. However, in visual image reconstruction with high-dimensional output space, evaluating zero-shot prediction is essential. Models must demonstrate true generalization to unseen stimuli rather than reflecting training data biases for establishing both reliability and practical utility.}

One of the remaining challenges in visual image reconstruction is the development 
of metrics for evaluating the quality and accuracy of the reconstructed images. 
The first and most critical step in assessing reconstruction results is to 
confirm a qualitative similarity between the reconstructed images and the 
perceived images through visual inspection across a diverse range of test sets. 
Following this, quantitative metrics should be employed for a more objective, 
high-throughput evaluation. However, as our analysis has suggested, it can be 
misleading to evaluate reconstruction by heavily relying on identification 
performance based on the relative similarity among alternatives \citep{koide-majima_mental_2024}. 
Even in cases where the reconstructed images only capture superficial 
information, such as categories or overall brightness, identification metrics 
can still be high. While identification performance can provide a useful 
benchmark, it should not be the sole metric for evaluating reconstruction quality. 
It is crucial to develop more appropriate similarity metrics that can accurately 
measure the perceptual similarity between the reconstructed and original images. 
One promising approach is to leverage image quality assessment (IQA) techniques 
from the computer vision field \citep{fu_dreamsim_2023,ding_image_2020}. 
These techniques are designed to quantify the perceptual quality of images and
can be adapted to the specific requirements of visual image reconstruction. 

Our findings have profound implications for research integrity and responsible dissemination of scientific results. Visual image reconstruction methods have gained attention not only from 
neuroscientists but also from the general public and policymakers, sparking 
discussions about their potential applications and risks \citep{unesco_unveiling_2023}. 
These stakeholders often contemplate the possibilities of seamless information 
communication through the brain, such as in brain-machine interfaces (BMIs), 
or the dangers of unauthorized access to private information from brain activity. 
This interest may stem from the perception that brain activity data can be 
obtained easily and reliably in real-time. However, current technology and 
analysis methods fall short of these expectations. Beyond the limitations discussed in this study, there are additional challenges in the field. Most 
reconstruction methods analyze previously acquired brain data offline. 
The brain data used for reconstructing images are often averaged 
over multiple presentations of the test image, with only a few studies 
demonstrating single-trial reconstruction results \citep{miyawaki_visual_2008,cheng_reconstructing_2023}. 
Further, it has been argued that subject cooperation is essential for reliably 
training and testing decoding models \citep{tang_semantic_2023}. Given these realities, public expectations often exceed current capabilities, and meeting these high demands in the short term is challenging. By clearly articulating these constraints, we can help manage expectations, prevent disappointment, and guide governments and companies away from misguided decisions. It is crucial to resist making overly optimistic claims about the ability to reconstruct arbitrary images. In light of these challenges, while the field of visual image reconstruction from brain activity holds great promise, it is our responsibility as researchers to ensure that its current capabilities and limitations are accurately communicated to all stakeholders.

\subsection{Recommendations}

\rev{
Finally, we present several guidelines for critically testing visual image reconstruction methods. These suggestions build upon the limitations and challenges identified in our study and provide a pathway for future improvements in reconstruction research.
}

\subsubsection{Stimulus design and data splits}
\rev{
\textbf{Expand and control diversity.}  Collect or curate training datasets that span sufficient axes in the feature space so that new (unseen) visual images can be predicted. When possible, include artificial or carefully-designed stimuli as well as natural stimuli in the test set to provide clearer interpretability and control. Natural image reconstruction is not necessarily the ultimate goal, given that humans perceive both natural and artificial images.
}

\rev{
\textbf{Avoid overlaps in training and test stimuli.} To evaluate true zero-shot capacity, ensure that test images do not overlap semantically or visually with training images. Identifying and removing near-duplicates or highly similar images from the test set helps prevent hidden ``shortcut'' solutions in which the model simply memorizes or classifies into known stimuli.
}

\rev{
\textbf{Use multiple and independent test sets.} Consider separate test sets with varying complexity—\textit{e.g.}, natural images, artificial shapes, and out-of-distribution samples—to comprehensively assess generalizability. Disentangling performance across these varied sets can reveal whether a method is genuinely reconstructing novel content or only handling a narrow range of stimuli.
}

\subsubsection{Model specification and latent feature choice}
\rev{
\textbf{Confirm the generator’s recovery capability.} Perform ``recovery checks'' by feeding true latent features (extracted directly from the original images) into the image generator. If the generator fails to reproduce the original images faithfully, it cannot serve as a valid reconstruction module. This step clarifies whether errors in the test phase stem from the latent translator or from a generator prone to hallucinations.
}

\rev{
\textbf{Use compositional, image-preserving features.} Favor latent features (such as mid-level DNN layers) that retain sufficient image-level detail and compositional representations. This ensures that, with perfect translation from the brain, the original image can be reconstructed accurately. In contrast, purely semantic or text-based features often discard important visual details, limiting reconstruction fidelity.
}

\rev{
\textbf{Mitigate output dimension collapse.} Choose or design translators (\textit{e.g.}, modular or sparse voxel selection approaches) to reduce collapse onto limited training-set clusters. Avoid overfitting to narrow categories by ensuring that the model maintains high-dimensional predictive capacity across all important feature dimensions.
}

\subsubsection{Evaluation metrics and result transparency}

\rev{
\textbf{Prioritize perceptual resemblance across diverse targets.} Before focusing on quantitative metrics, visually confirm that reconstructions capture the perceptual features of all target stimuli. Testing fidelity on diverse or out-of-distribution samples is crucial for confirming genuine reconstruction rather than mere classification or retrieval. Avoid overemphasizing photorealism, as it can mask inaccuracies. Include extensive examples so readers can visually assess quality and variation.
}

\rev{
\textbf{Avoid cherry-picking and post-hoc selection.} Generative models can produce multiple plausible outputs from a single latent feature. Selecting only the best-looking or most accurate images artificially inflates performance estimates. Present results transparently (\textit{e.g.}, showing random draws or evaluating robustness across different seeds) to offer a fair depiction of each model’s true reliability.
}

\rev{
\textbf{Use robust metrics beyond pairwise identification.} Pairwise identification is easy to implement, but can overestimate performance especially in categorically structured data. Carefully design the selection of candidates when conducting identification analysis. In addition, supplement it with more stringent evaluation, such as fine-grained semantic checks, and advanced image-quality metrics (\textit{e.g.}, SSIM, DreamSim, other learned perceptual measures). Distinguish clearly between high-level semantic alignment and perceptual similarity.
}

\subsubsection{Collaboration and ethical communication}

\rev{
\textbf{Interdisciplinary collaborations.} Close collaboration among neuroscientists, machine learning researchers, and cognitive scientists is vital for designing robust experiments, interpreting results correctly, and addressing complex technical pitfalls (\textit{e.g.}, data leakage, improper splits, or hallucinations by diffusion models).
}

\rev{
\textbf{Transparent reporting and data-sharing.} Provide open-source code, clearly document training--test splits, and release relevant stimuli annotations to enable reproducibility. Transparency fosters collective progress, allowing others to replicate or extend your findings under more controlled or diverse conditions.
}

\rev{
\textbf{Realistic public and policy discourse.} Communicate clearly that current reconstruction methods do not equate to unconstrained “mind reading” and often depend on carefully curated data. Highlight the role of subject cooperation, offline averaging, and limited generalizability so that stakeholders—such as policymakers, journalists, and the public—avoid overestimating immediate real-world capabilities.
}

\rev{
In sum, authentic visual image reconstruction from brain activity requires careful management of dataset diversity and overlap, prudent model specification (especially in latent feature selection), and rigorous evaluation metrics beyond simple identification. By adhering to these recommendations, researchers can reduce the risk of reporting spurious reconstructions, bringing us closer to methods that genuinely reflect an individual’s perceptual experience. 
}

\section{Conclusions}
Our critical analysis of recent generative AI-based visual image reconstruction methods revealed several limitations and challenges. We demonstrated that the apparent success of text-guided reconstruction methods primarily stems from a combination of classification into trained categories and hallucination through text-to-image diffusion, rather than genuine zero-shot reconstruction of novel images, which was the original goal of reconstruction studies. Our formal analysis revealed that predicting features with limited diversity can lead to output dimension collapse, where predictions become confined to patterns similar to the training set. Our simulation analysis demonstrated that successful zero-shot prediction requires training data with sufficient diversity to span the effective dimensions of the target feature space. We also pointed out that standard identification metrics can be misleading, especially when the target set has an underlying similarity structure. Additionally, we provided evidence that much of the input information is preserved at almost all hierarchical levels of deep neural networks. Finally, we pointed out that recent realistic reconstructions produced by generative AI models may appear convincing but do not necessarily reflect accurate representations of perceived visual experiences. These findings emphasize the need for more rigorous evaluation methods, diverse datasets, and careful interpretation of results in visual image reconstruction research. Future work should focus on prioritizing accurate reconstruction rather than naturalistic appearance. This objective would be achieved by utilizing compositional representations that can effectively span the feature space while maintaining the ability to recover original stimuli from latent features. As the field continues to attract attention from both researchers and the public, our results have important implications for research integrity and responsible development of neurotechnology, highlighting the need to balance scientific advancement with realistic expectations about current technological capabilities.





\section*{Methods}

\subsection*{Datasets}
We utilized two datasets: the Natural Scene Dataset (NSD; \citealp{allen_massive_2021}) 
and the Deeprecon dataset \citep{shen_deep_2019}. Both datasets comprise visual 
stimuli and corresponding fMRI activity collected when subjects perceived the 
stimuli. In the NSD dataset, eight subjects were presented with MSCOCO 
images \citep{lin_microsoft_2014}, yielding $30,000$ brain activity samples per 
subject, which is three times the amount provided by the Deeprecon dataset. 
The Deeprecon dataset includes fMRI activity data from subjects presented 
with both ImageNet images \citep{deng_imagenet:_2009} and artificial images. 
It contains roughly $8,000$ brain samples per subject. Since this dataset is 
designed to evaluate reconstruction performance, the test stimuli were carefully 
selected. The test natural images were selected from ImageNet, which were in 
categories different from those used in the training. The artificial images 
were only used as test data to check the generalizability performance of 
the proposed reconstruction methods. In both datasets, we adopted the training--test 
split used in previous studies and utilized data from the first subject 
(S1 in the NSD and  Subject 1 in the Deeprecon). Text-guided reconstruction 
methods require text annotations of images. For the NSD, text annotations 
accompanying the MSCOCO database were used. For the Deeprecon dataset, 
we collected text annotations for each experimental stimulus via crowd workers on 
Amazon Mechanical Turk, yielding five annotations per image. The text annotations of 
training stimuli are publicly available in the GitHub repository 
(\url{https://github.com/KamitaniLab/GOD_stimuli_annotations}).

\subsection*{Reconstruction methods}

We utilized three image reconstruction methods: StableDiffusionReconstruction \citep{takagi_high-resolution_2023}, 
Brain-Diffuser \citep{ozcelik_natural_2023}, \rev{MindEye2 \citep{scotti_mindeye2_2024}}, and iCNN \citep{shen_deep_2019}. Each 
method employs two common steps: first, translating brain activity patterns 
into latent features of the stimuli, and second, generating images from these 
latent features using an image generator (Fig.~\ref{Figure_01}). In the 
StableDiffusionReconstruction method, the latent features are the VAE \citep{kingma_auto-encoding_2014}
features calculated from stimulus images and the CLIP text features \citep{radford_learning_2021} 
from the image annotations. The generator is StableDiffusion \citep{rombach_high-resolution_2022}. 
They first generate low-resolution images from the translated VAE features, 
and those images are further fed into the StableDiffusion model with translated 
text features to generate images. The generated images are regarded as 
reconstructed images from brain activity. In the Brain-Diffuser method, the 
latent features are the VDVAE \citep{child_very_2021}, CLIP vision features 
from stimulus images and CLIP text features from the text annotations of the image. The 
generator is Versatile diffusion \citep{xu_versatile_2022}. Similar to 
StableDiffusionReconstruction, low-resolution images are first generated from 
the translated VDVAE features, and these images are further used for the input 
of the versatile diffusion model with the translated vision and text features.
The generated images are regarded as reconstructed images from brain activity. 
\rev{In the MindEye2 methods, the latent features are the variant of CLIP vision model features (OpenCLIP ViT/bigG-14). The generator is multiple Stable diffusion XL (SDXL) models \citep{podell_sdxl_2023}. They first generate images from translated OpenCLIP ViT/bigG-14 features by unCLIP technique \citep{ramesh_hierarchical_2022, scotti_mindeye2_2024}. They then generate the final reconstruction by SDXL, integrating the generated images and text caption predicted from the translated OpenCLIP ViT/bigG-14 features by GiT Image2Text modules \citep{wang2022git}.}
In the iCNN method, the latent features are the intermediate output of the VGG19 layer \citep{simonyan_very_2014}. As a generator, they used the pre-trained image generator \citep{dosovitskiy_generating_2016}, 
and they solved the optimization problem to minimize the discrepancy between 
the VGG19 features calculated from the generated images and the translated 
VGG19 features. Well-optimized images are regarded as reconstructed images. 

\subsection*{UMAP visualization}

To investigate dataset diversity, we employed uniform manifold approximation 
and projection (UMAP), a nonlinear dimensionality reduction technique \citep{mcinnes_umap:_2018}, 
to learn a projection from a latent features space to a lower dimension 
(UMAP embedding space). We used both the training and test CLIP text features 
to learn the UMAP projection. These features were combined and standardized 
beforehand. The hyperparameters followed the official guide for 
clustering usage (\href{https://umap-learn.readthedocs.io/en/latest/clustering.html#umap-enhanced-clustering}{https://umap-learn.readthedocs.io/en/latest/clustering.html})
with cosine distance as a distance metric. The learned UMAP was also used to project the features 
predicted from brain activity (Fig.~\ref{Figure_06}B). After standardizing 
the predicted features using the same mean and standard deviation parameters 
used in UMAP projection learning, we projected the predicted features onto 
the UMAP embedding space. 

\subsection*{Simulation with clustered data}

We conducted a simulation analysis to illustrate output dimension collapse 
(Fig.~\ref{Figure_08}) and to examine the generalization performance beyond 
the training data (Fig.~\ref{Figure_09}). These analyses involve a 
teacher-student learning task. The teacher model generated the pairs of target features and brain activity as training samples, and the student model learned a mapping from the training samples. To imitate the feature translation situation from brain activity, observation noise was added to the brain data.

The training sample of latent feature data $\mathbf{y} \in \mathbb{R}^D$ was generated from a Gaussian mixture distribution, formulated as:
\begin{equation}
  p_\mathrm{tr}(\mathbf{y}) = \frac{1}{C} \sum_{c=1}^{C} \mathcal{N}(\boldsymbol{\mu}_c^\mathrm{ tr},\sigma_{\mathrm{intra}}^2I), \quad \boldsymbol{ \mu}_c^\mathrm{ tr} \sim \mathcal{ N}(\mathbf{ 0}, \sigma_{\mathrm {inter}}^2I),
\end{equation}
where $C$ is the number of clusters in the training set. $\sigma_{\mathrm{intra}}^2$ is the scalar value representing the 
variance of the Gaussian distribution corresponding to each cluster. $\sigma_{\mathrm {inter}}^2$ is the scalar value representing 
the variance of the distribution of cluster centers ($\boldsymbol{\mu}_c^\mathrm{tr}$). $I$ is the $D\times D$ identity matrix. Brain activity data, 
$\mathbf{x} \in \mathbb R^D$, were created using teacher weights $\bar A\in \mathbb R^{D\times D}$ and incorporating observation noise $\boldsymbol{\xi}$ with 
$\mathbf{ x} = \bar A^{\top}\mathbf{ y} + \boldsymbol{\xi}$, where $\boldsymbol{\xi} \sim \mathcal N(\mathbf{0},\sigma_{\mathrm {noise}}^2 I)$. $\sigma^2_{\mathrm {noise}}$ 
is the scalar value representing the variance of observation noise. 

For $N$ training samples, $X_{\mathrm{tr}}=[\mathbf{x}^{(1)}, \ldots, \mathbf{x}^{(N)}]^\top$ and $Y_{\mathrm{tr}}=[\mathbf{y}^{(1)}, \ldots, \mathbf{y}^{(N)}]^\top$, we trained the student model. by the ridge regression algorithm. The trained weight of ridge regression model $W$ can be calculated analytically:
\begin{equation}
	W =(X_\mathrm{tr}^{\top}X_\mathrm{tr} + \lambda I)^{-1}X_\mathrm{tr}^{\top} Y_\mathrm{tr},
\end{equation}
where $\lambda$ is the regularization parameter.

After training the student weight $W$, we illustrated the phenomenon of output dimension collapse as in Fig.~\ref{Figure_08} by predicting randomly generated data. First, we generated 
random test latent features from the Gaussian distribution with 
a mean of $\mathbf{0}$ and a variance equal to the training set variance scaled by a 
single scalar value ($9.0$). The corresponding test brain samples were 
obtained by translating the latent features by the teacher's weight 
$\bar A$ and adding observation noise $\boldsymbol{\xi}$. The predicted latent features 
were derived by projecting the test brain samples by the learned student weight $W$. 
We set $D = 512$, $\sigma_{\mathrm{intra}}^2 = 10/512$, $\sigma_{\mathrm{inter}}^2 = 100/512$, and $\sigma_{\mathrm{noise}}^2 = 10$.

We evaluated the zero-shot performance of the student model as in Fig.~\ref{Figure_09}. We prepared two types of test samples: in-distribution test samples and out-of-distribution 
test samples. In-distribution test samples are generated from one of the (Gaussian) clusters used in training data:
\begin{equation}
  p_\mathrm{{te}}(\mathbf{y}) =\mathcal N(\boldsymbol{\mu}^\mathrm{tr}_{c},\sigma_{\mathrm{intra}}^2I),
\end{equation}
where we randomly chose $\boldsymbol{\mu}_c^{\mathrm{tr}}$ over $C$ cluster centers.
Out-of-distribution (OOD) test samples were generated from a novel cluster center $\boldsymbol{\mu}^\mathrm{ood}$ that is not included in the training set. The novel cluster center was obtained by sampling from a Gaussian distribution $\boldsymbol{\mu}^\mathrm{ood}\sim \mathcal{ N}(\mathbf{ 0}, \sigma_{\mathrm{inter}}^2I)$. 
The OOD test samples were then generated from the novel cluster center:
\begin{equation}
  p_\mathrm{{ood}}(\mathbf{ y}) =\mathcal N(\boldsymbol{ \mu}^\mathrm{ ood},\sigma_{\mathrm{intra}}^2I).
\end{equation}

To evaluate the model's zero-shot performance, we conducted a cluster 
identification analysis. For each test sample, we calculated the 
similarity between its predicted features and the center of its original cluster, as well as the 
similarity between its predicted features and the centers of the other candidate clusters. 
Each test sample was assigned to the cluster whose center had the highest similarity with its predicted features, and the proportion of samples correctly assigned to their true cluster centers was calculated across all $n$ test samples. We used the correlation coefficient as the similarity measure. To reduce the variability associated with the cluster selection, 
we repeated this process $t$ times by randomly selecting both in-distribution and out-of-distribution cluster centers and reported the median value of the results. For in-distribution test samples, we chose the cluster center randomly without replacement.

We mainly explored the dependency of the typical cluster identification performance on the following hyperparameters: dimension $D$, the number of clusters 
in the training set $C$, and the ratio of variance about the cluster structure $\sigma_{\mathrm {intra}}^2/\sigma_{\mathrm {inter}}^2$. 
We used a sufficiently large number of training samples $N = 500,000$, and we kept $N$ constant while changing the above hyperparameters, especially $C$. Other parameters were also fixed in this simulation as follows: 
$\sigma_{\mathrm{intra}}^2 + \sigma_{\mathrm{inter}}^2 = 110/D$, ${\bar A_{ij}} \sim\mathcal{N}(0,D^{-1/2})$, $\sigma_{\mathrm {noise}}^2 = 0.25$, 
$\lambda=1.0$, $n = 100$ and $t=32$. We parameterized the scale of the variances ($\sigma_{\mathrm {intra}}^2$ and $\sigma_{\mathrm {inter}}^2$) and the teacher weights {$\bar A_{ij}$} using the dimension $D$ so that the order of 
scale of the output samples were invariant from the dimension. When the number of clusters in the training set $C$ 
was less than $t=32$, we set $t=C$ instead.

\subsection*{Expected identification accuracy in imprecise reconstructions}

A pairwise identification accuracy is a metric defined on three types of samples: the test sample, the predicted 
sample, and the candidate sample selected from a test set as 
$\mathbf{y}, \hat{\mathbf{y}}, \mathbf{y}_{-} \in \mathcal{Y}$, respectively. We define a function 
$S \colon \mathcal{Y} \times \mathcal{Y} \times \mathcal{Y} \rightarrow \mathbb{R}$ that takes the triplet above as the input and output whether the predicted sample was much closer to the test sample than the candidate 
sample as
\begin{equation}
	S(\hat{\mathbf{y}}, \mathbf{y}, \mathbf{y}_{-}) = \begin{cases}
		1 & \left( \operatorname{sim}(\hat{\mathbf{y}}, \mathbf{y}) \ge \operatorname{sim}(\hat{\mathbf{y}}, \mathbf{y}_{-}) \right) \\
		0 & \left( \text{otherwise} \right)
	\end{cases}
\end{equation}
where $\operatorname{sim}(\cdot, \cdot)$ is an arbitrary function that evaluates a similarity between two samples. 
The pairwise identification accuracy $\mathrm{Acc}$ over $n$ test samples is defined as
\begin{equation}
	\mathrm {Acc} = \frac{1}{n(n-1)}\sum_i^n\sum_{j\neq i}^n S(\hat{\mathbf{y}}^{(i)}, \mathbf{y}^{(i)}, \mathbf{y}_{-}^{(j)}).
\end{equation}
Now, we consider a scenario where the translator only decodes semantic information (\textit{e.g.}, category) and cannot 
decode information about its precise visual appearance. Suppose the test set contains a categorical structure like
NSD stimuli, we model such a scenario as
\begin{align}
	\mathbb{E}_{p(\mathbf{y}, \hat{\mathbf{y}}, \mathbf{y}_{-} \mid (\mathbf{y}, \hat{\mathbf{y}}, \mathbf{y}_{-}) \in Z)} [S(\hat{\mathbf{y}}, \mathbf{y}, \mathbf{y}_{-})] &= 0.5, \\
	\mathbb{E}_{p(\mathbf{y}, \hat{\mathbf{y}}, \mathbf{y}_{-} \mid (\mathbf{y}, \hat{\mathbf{y}}, \mathbf{y}_{-}) \in \bar{Z})} [S(\hat{\mathbf{y}}, \mathbf{y}, \mathbf{y}_{-})] &= q \text{ where } q \in [0.5, 1].
\end{align}
$Z$ is a set of triplets in which the test sample and the candidate sample belong to the same category. $\bar{Z}$ 
is a complementary set of $Z$. $\mathbb{E}_{p(\mathbf{y}, \hat{\mathbf{y}}, \mathbf{y}_{-} \mid \cdot)} 
[S(\hat{\mathbf{y}}, \mathbf{y}, \mathbf{y}_{-})]$ represents the pairwise identification accuracy in the 
conditional expectation form. If the candidate sample belongs to the same category as the test sample, pairwise 
identification is challenging because of the poor prediction of the translator. On the other hand, if the candidate 
samples belong to a different category than the test sample, the test sample is easily identified only from the 
semantic information.

Here, we assume that the test set contains $k$ categories in total and that all samples are equally distributed 
across each category for simplicity. If we have a sufficiently large number of test samples, the above 
identification accuracy can be approximated as
\begin{align}
	\mathrm {Acc} &\underset{\phantom{n \to \infty}}{=} \frac{1}{n(n-1)}\left(\sum_{(\mathbf{y}, \hat{\mathbf{ y}}, \mathbf{y}_{-}) \in Z} S(\hat{\mathbf{y}}, \mathbf{y}, \mathbf{y}_{-})\right) \notag \\
        &\hspace{3.5cm}+ \frac{1}{n(n-1)}\left(\sum_{(\mathbf{y}, \hat{\mathbf{y}}, \mathbf{y}_{-}) \in \bar Z} S(\hat{\mathbf{y}}, \mathbf{y}, \mathbf{y}_{-})\right)  \\
	&\underset{\phantom{n \to \infty}}{=} \frac{|Z|}{n(n-1)}\left(\frac{1}{|Z|}\sum_{(\mathbf{y}, \hat{\mathbf{y}}, \mathbf{y}_{-}) \in Z} S(\hat{\mathbf{y}}, \mathbf{y}, \mathbf{y}_{-})\right) \notag \\
    &\hspace{3.5cm} + \frac{|\bar Z|}{n(n-1)}\left(\frac{1}{|\bar Z|}\sum_{(\mathbf{y}, \hat{\mathbf{y}}, \mathbf{y}_{-}) \in \bar Z} S(\hat{\mathbf{y}}, \mathbf{y}, \mathbf{y}_{-})\right) \\
	&\underset{n \to \infty}{=} \frac{1}{k} \cdot \mathbb{ E}_{p(\hat{\mathbf{y}},\mathbf{y}, \mathbf{y}_{-} \mid (\mathbf{y}, \hat{\mathbf{y}}, \mathbf{y}_{-}) \in Z)} [S] \notag \\
    &\hspace{3.5cm} + \left(1 - \frac{1}{k}\right) \cdot \mathbb{ E}_{p(\hat{\mathbf{ y}}, \mathbf{y}, \mathbf{y}_{-} \mid (\mathbf{y}, \hat{\mathbf{y}}, \mathbf{y}_{-}) \in \bar{Z})} [S] \\
	&\underset{\phantom{n \to \infty}}{=} \frac{1}{k}\cdot0.5 + \left( 1-\frac{1}{k} \right) \cdot q,
\end{align}
where $|Z| = n\left(n/k-1\right)$, and $|\bar Z| = n\left(n-n/k\right)$.
We used the assumption of a large sample size at the third equality.

\subsection*{Recovery check of a single layer by iCNN}

We performed a recovery check analysis using a single layer from the iCNN method in Fig.~\ref{Figure_11}. The iCNN method generates an image by optimizing pixel values to 
make the image's latent features similar to the target latent features 
\citep{shen_deep_2019}. In the pixel optimization condition (the left columns 
of each recovery image in Fig.~\ref{Figure_11}), we directly optimized the 
pixel values of images to minimize the mean squared loss between the latent 
features of the image as well as the total-variance (TV) loss of pixel values 
\citep{mahendran_understanding_2015}.

Additionally, the iCNN method can incorporate image generator networks
(middle and right columns of each recovery image in Fig.~\ref{Figure_11}) 
to add constraints on image statistics. Instead of optimizing the pixel values, 
we optimized the parameters related to the generator networks to minimize the 
mean squared loss between the latent features obtained through the generator 
networks and the target latent features. As a weak image prior, we used Deep 
Image Prior (DIP; \citealp{ulyanov_deep_2017}). DIP utilizes a hierarchical U-Net 
architecture as an inherent prior for image tasks, capturing the statistical 
regularities of images without relying on a specific dataset. This model 
works effectively by optimizing a randomly initialized neural network 
that can be used as an image prior in various inverse problems such as 
denoising, super-resolution, and inpainting tasks. In our analysis, DIP 
started with a U-Net initialized with random noise. Subsequently, the latent 
features and weight parameters of DIP were optimized to minimize the difference 
between the network's output and target DNN features. For the pre-trained image 
prior \citep{dosovitskiy_generating_2016}, we used the same generator model 
as in \cite{shen_deep_2019} optimizing the latent features of 
the pre-trained networks.

\section*{Acknowledgments}
We thank our laboratory team, especially Eizaburo Doi, Haibao Wang, Kenya Otsuka, Hideki Izumi, and Matthias Mildenberger, for their invaluable feedback and insightful suggestions on the manuscript.





\clearpage
\appendix
\renewcommand{\thefigure}{A\arabic{figure}}
\setcounter{figure}{0}
\section{Supplementary figures}

\begin{figure}[!h]
  \centering
    \includegraphics[width=1.0\linewidth]{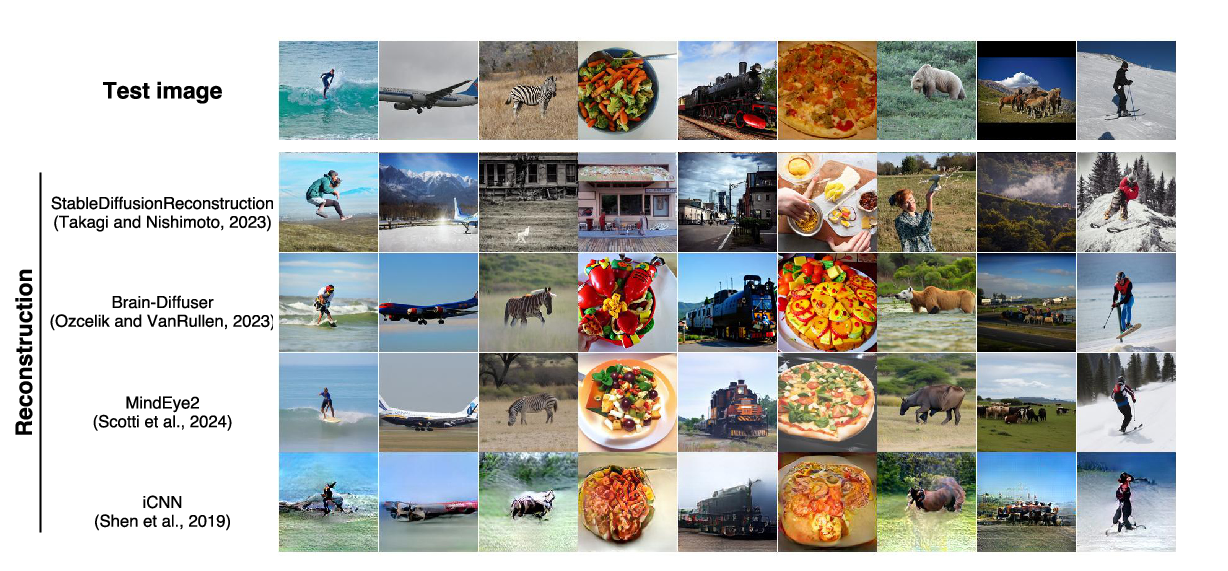}

  \caption{\textbf{Reconstructions from the NSD dataset using a sample size matched to the deeprecon dataset.} 
  \rev{The figure follows the format of Fig.~\ref{Figure_02} in the main text. The text-guided reconstruction and MindEye2 methods did not show a large performance drop, even when the training sample size of the NSD dataset was reduced to match that of the Deeprecon dataset. This result supports that the degraded performance in the Deeprecon is not simply due to its smaller sample size compared to the NSD.}}
  \label{Figure_A1}
\end{figure}

\begin{figure}[!b]
  \centering
    \includegraphics[width=1.0\linewidth]{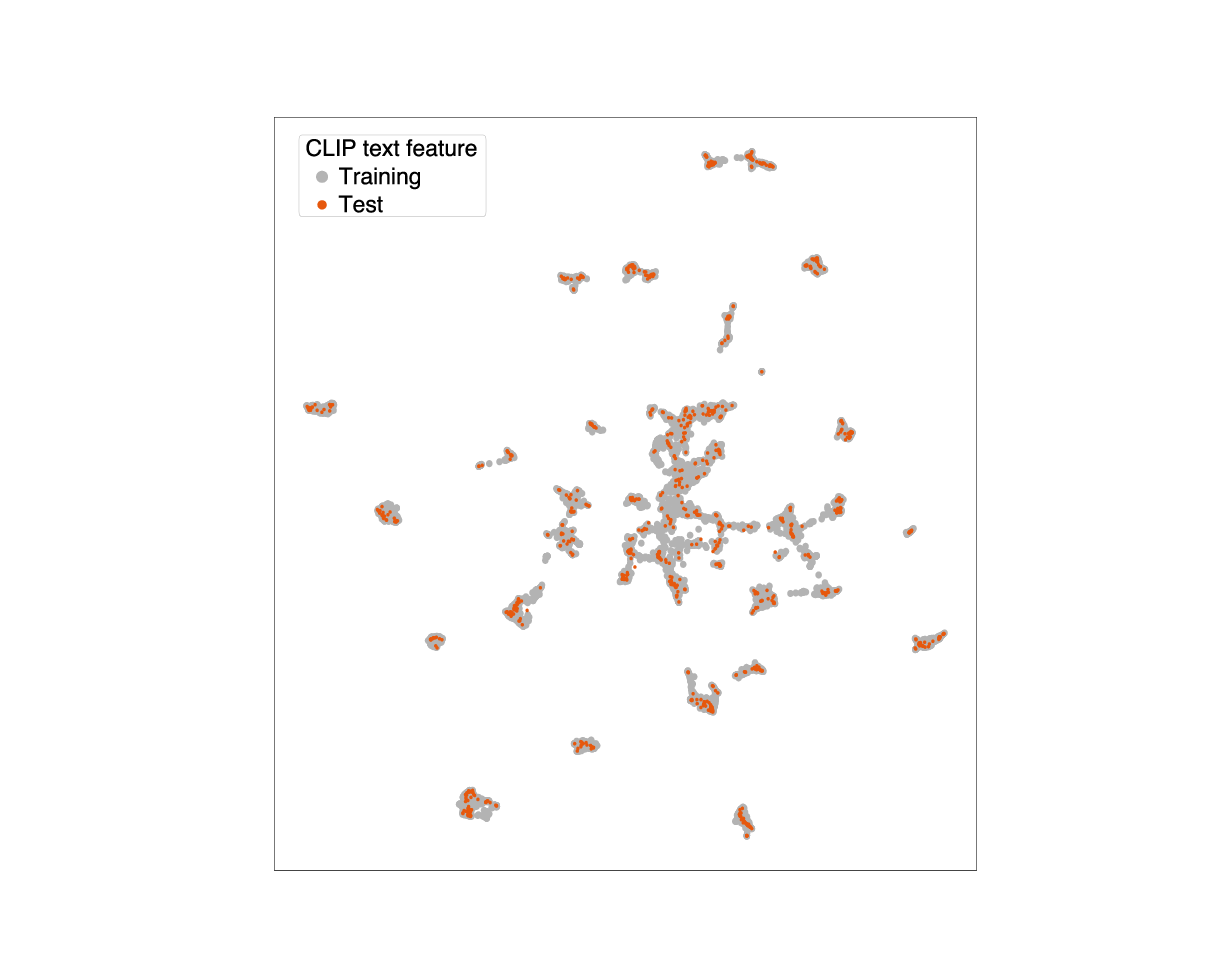}

  \caption{\textbf{UMAP visualization of CLIP text features in the Natural Scene Dataset (NSD) using default parameters.} This scatter plot, analogous to Fig.~\ref{Figure_04}A in the main text, depicts the distribution of text features within the NSD. The gray points represent training samples, and the orange points represent test samples. Unlike Fig.~\ref{Figure_04}A, which used the parameters optimized for clustering visualization, this figure employs the default UMAP settings.  Despite the absence of parameter tuning,  the plot still reveals discernible clusters and considerable overlap between training and test samples, indicating that the observed clustering pattern does not hinge on specialized UMAP configurations.
}
  \label{Figure_A2}
\end{figure}

\clearpage
\begin{figure}[!t]
  \centering
    \includegraphics[width=1.0\linewidth]{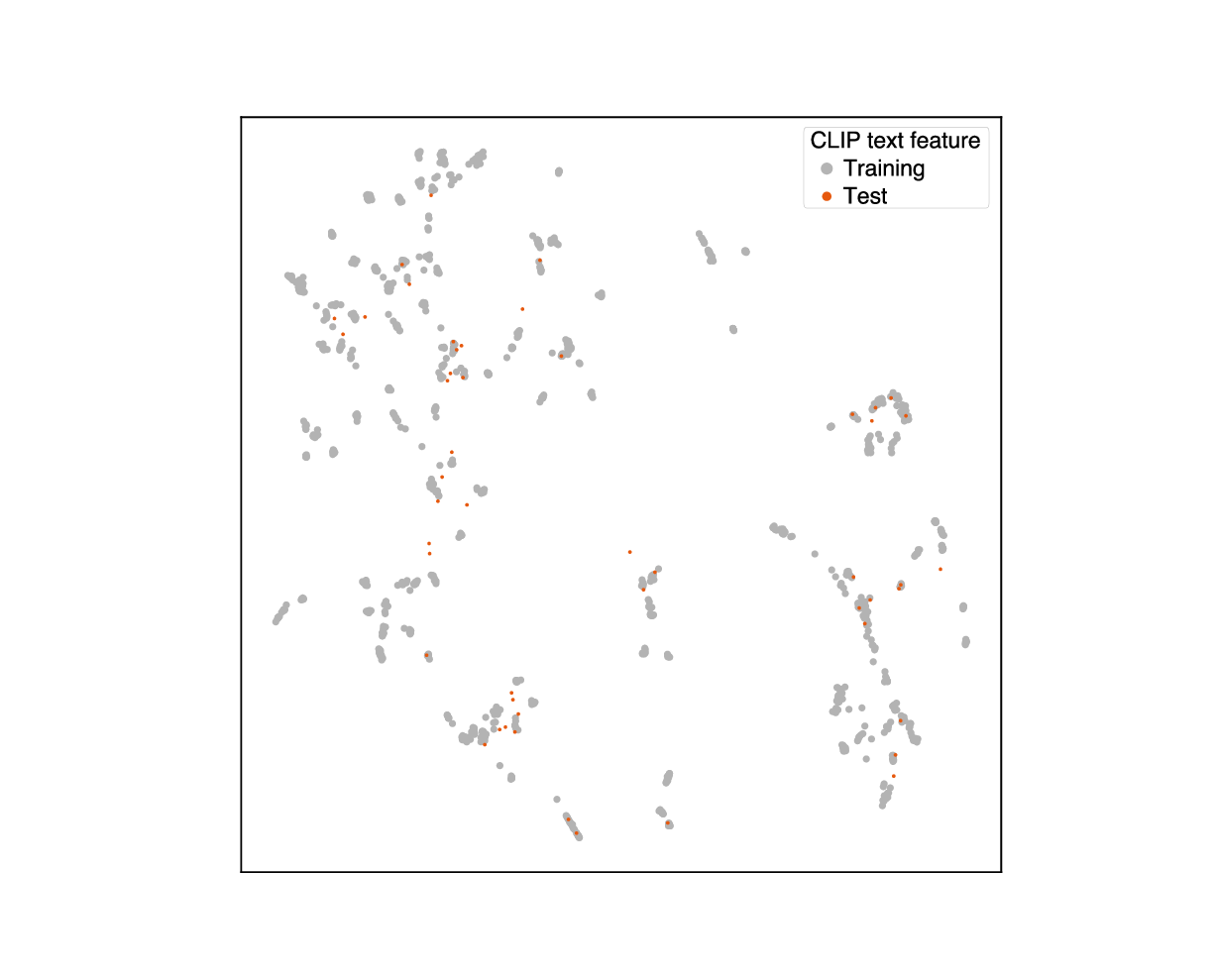}

  \caption{\textbf{UMAP visualization of CLIP text features in the Deeprecon dataset.} This scatter plot visualizes the distribution of semantic features within the Deeprecon data, with the gray points representing training samples and the orange points indicating test samples. Unlike the NSD dataset (Fig.~\ref{Figure_04}A in the main text), this visualization demonstrates a clearer separation between training and test samples. This distinction highlights the Deeprecon dataset's intentional design to differentiate object categories between training and test sets.}

\label{Figure_A3}
\end{figure}

\clearpage
\begin{figure}[!t]
  \centering
    \includegraphics[width=1.0\linewidth]{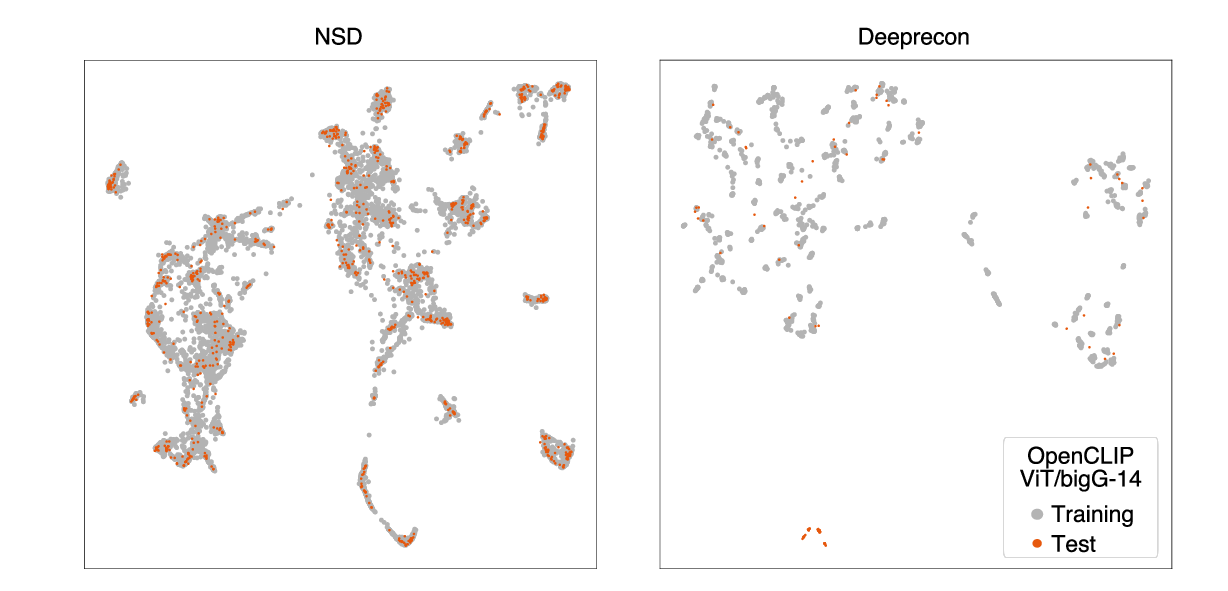}

  \caption{\textbf{UMAP visualization of the latent features of MindEye2 (OpenCLIP ViT/bigG-14) in the NSD and Deeprecon datasets.} \rev{The plots follow the format of Fig.~\ref{Figure_04}A and Fig.~\ref{Figure_A3}. While these features are extracted from images, not from text captions, the plot still reveals a cluster structure, with considerable overlap between the training and test sets in the NSD  dataset, whereas the Deeprecon dataset exhibits noticeably less overlap.}}

\label{Figure_A4}
\end{figure}

\clearpage
\begin{figure}[!t]
  \centering
    \includegraphics[width=1.0\linewidth]{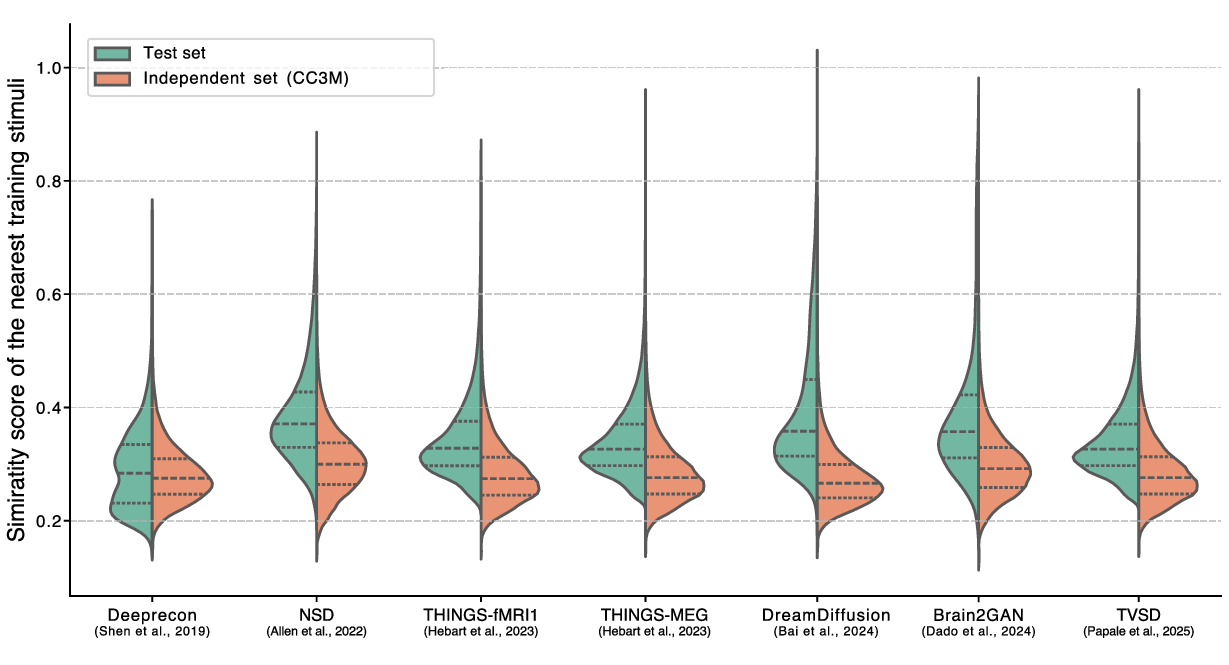}

  \caption{\textbf{Distribution of similarity scores between training and test stimuli.} \rev{Violin plots show the distribution of DreamSim-based similarity scores \citep{fu_dreamsim_2023} between each dataset’s training set and its test stimuli (green) compared to the similarity distributions between the training set and an independent dataset (orange; randomly selected 1000 images in CC3M; \citealp{sharma_conceptual_2018}). \revrev{For each dataset, we plotted the distribution of similarity scores from the top $5\%$ most similar training images to either test or independent set images. Results remained consistent when varying the percentage or number of selected training images.} A large deviation between the test- and independent-set distributions indicates that the test set is biased toward the training set, while smaller deviations suggest greater test set independence. This analysis was performed on seven datasets: Deeprecon ($1,200$ training and $90$ test images) and NSD (8,859 training and $982$ test images), and was further extended to THINGS-fMRI1 ($8,640$ training and $100$ test images; \citealp{hebart_things-data_2023}), THINGS-MEG ($22,248$ training and $200$ test images; \citealp{hebart_things-data_2023}), DreamDiffusion ($1,330$ training and $333$ test images; \citealp{bai_dreamdiffusion_2024}), Brain2GAN (4,000 training and 200 test images; \citealp{dado_brain2gan_2024}), and TVSD ($22,248$ training and $100$ test images; \citealp{papale_extensive_2025}). \revrev{Most datasets exhibited higher training-test similarity compared to training-independent similarity, suggesting these splits are not optimal for evaluating zero-shot prediction. In contrast, the Deeprecon dataset was specifically designed to exclude overlapping training categories, facilitating a more suitable evaluation of generalizability performance. The distributions of training-independent similarity were consistent across the datasets, suggesting comparable levels of representativeness among the datasets.}
  }
  }

\label{Figure_A5}
\end{figure}

\clearpage
\thispagestyle{empty}
\begin{figure}[!t]
  \vspace{-2cm}
  \centering
    \includegraphics[width=1.0\linewidth, trim=0 6mm 0 0, clip]{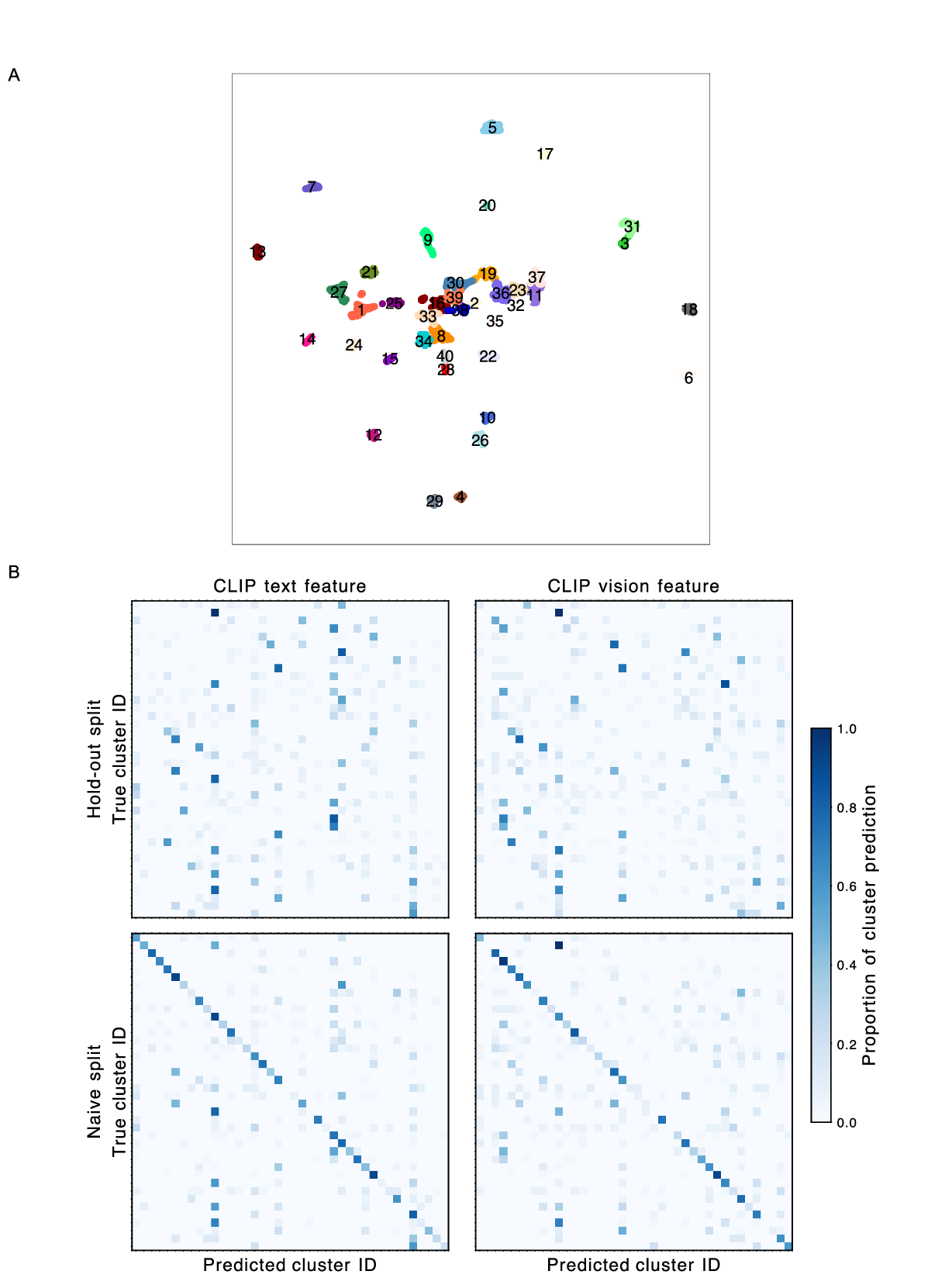}

  \caption{\textbf{Confusion matrices from the hold-out and naive splits.} (\textbf{A}) Clustering results of CLIP text features. The $k$-means clustering was applied to the UMAP embedding of CLIP text features from the NSD (Fig.~\ref{Figure_04}A in the main text). These clustering results were used to make a training--test split under the hold-out split condition (Fig.~\ref{Figure_06}A). 
(\textbf{B}) Confusion probability matrices for cluster identification. Left and right matrices represent CLIP text and vision features, respectively, with each cell $(i, j)$ indicating the proportion of samples from cluster $i$ predicted as cluster $j$. The top row shows results for the hold-out split, where entire semantic clusters are excluded from training, highlighting challenges in zero-shot prediction of novel categories. The bottom row shows the naive split, where semantic clusters appear in both training and test sets, demonstrating the translators' ability to identify learned semantic categories. 
}

\label{Figure_A6}
\end{figure}

\clearpage

\thispagestyle{empty}
\begin{figure}[t]
  \vspace{-3cm}
  \centering
    \includegraphics[width=0.8\linewidth]{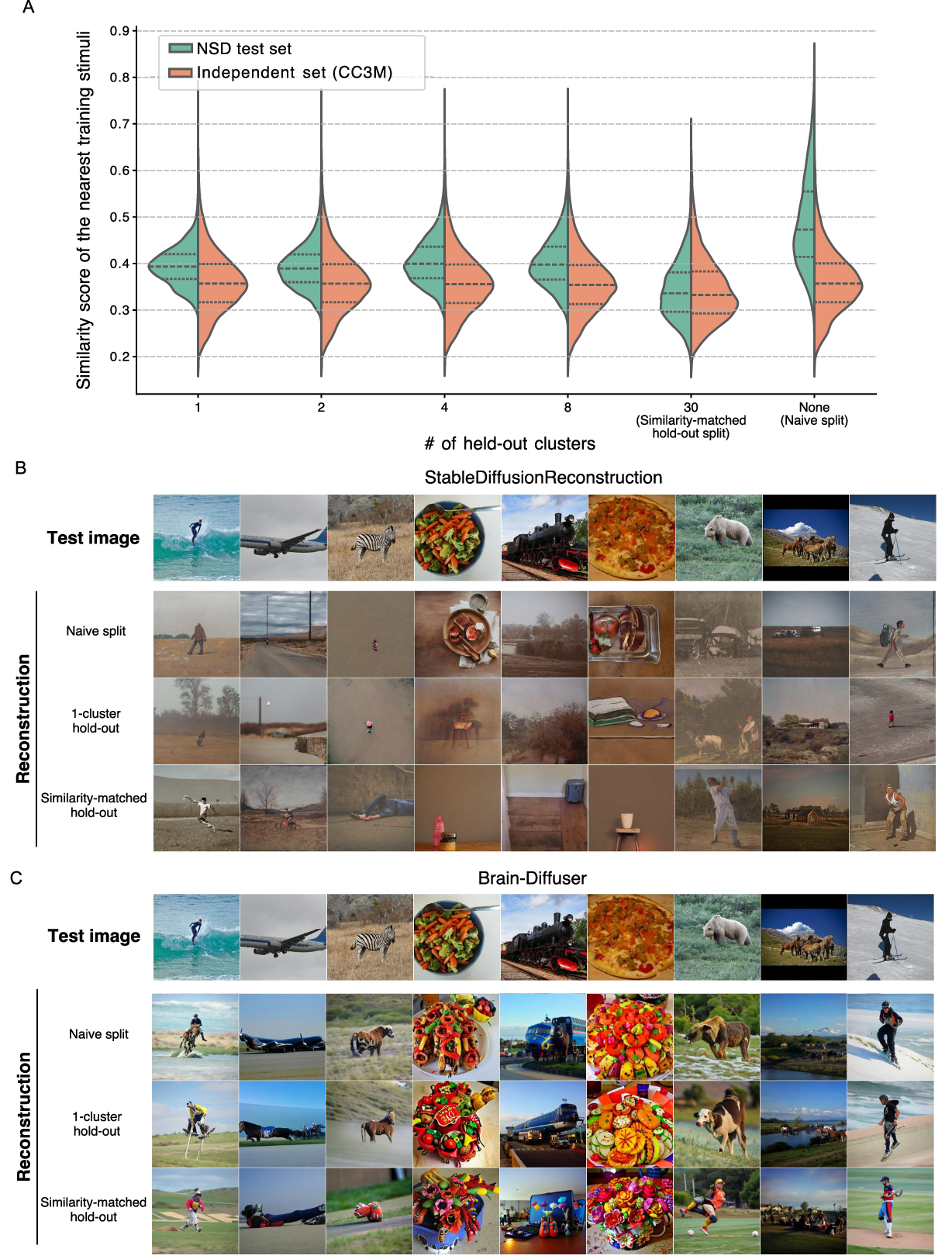}
  
  \caption{\textbf{Reconstruction analysis under hold-out split conditions using the NSD dataset.} \rev{Stimulus images were clustered via UMAP of CLIP text features into “held-out” and “training” groups; models were trained only on training clusters and evaluated on held-out clusters using independent brain data. (\textbf{A}) DreamSim similarity distributions for various numbers of held-out clusters. The figure shows similarities between the training set and the hold-out test set (green) versus those between the training set and an independent dataset (CC3M; orange) as in Fig.~\ref{Figure_A5}. The x-axis indicates the number of excluded clusters. Despite removals, the test set remains more similar to the training set than to CC3M. The ``similarity-matched hold-out'' condition aligns these similarities with Deeprecon (see Fig.~\ref{Figure_A5}). The ``naive split'' (right side of the dashed line) randomly removes samples from the original training set so that the total number of training samples matches that of the similarity-matched hold-out split. Unlike the similarity-matched hold-out split, the naive split allows cluster overlap between training and test sets. (\textbf{B}, \textbf{C}) Reconstructions under different split conditions for StableDiffusionReconstruction (\textbf{B}) and Brain-Diffuser (\textbf{C}). Each row shows the original test image (top), followed by reconstructions from the naive split, the 1-cluster hold-out, and the similarity-matched hold-out. All splits use the same number of training samples for fair comparison. For StableDiffusionReconstruction, regression parameters were adjusted to accommodate the smaller training set, yet the outputs remained low-quality. Although the naive-split results retain some semantic similarity to the test images, using a 1-cluster hold-out led to visibly and semantically different reconstructions from the original test images. Brain-Diffuser reconstructions reveal clearer differences across conditions: naive-split outputs closely resemble the test images, 1-cluster hold-out reconstructions preserve the overall layout with minor semantic changes, and similarity-matched hold-out reconstructions deviate substantially in both visual and semantic content.}}
\label{Figure_A7}
\end{figure}

\clearpage

\begin{figure}[!t]
  \centering
    \includegraphics[width=1.0\linewidth]{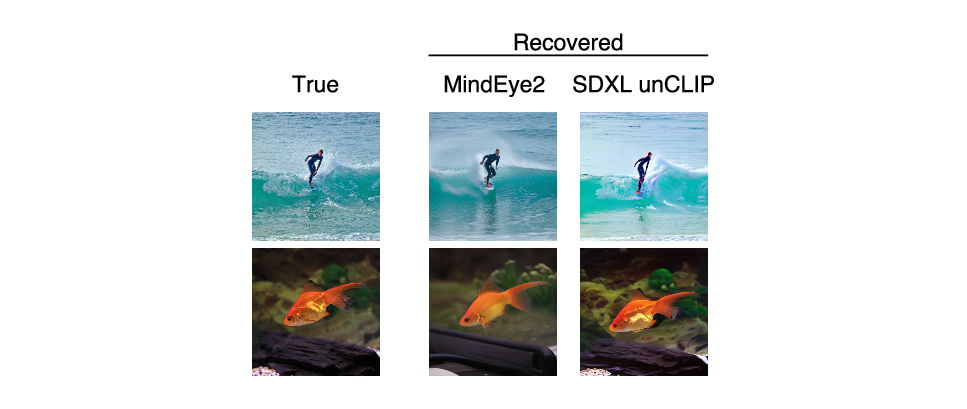}

  \caption{\textbf{Recovery check for MindEye2.} 
  \rev{The panels follow the format of Fig.~\ref{Figure_07} in the main text. MindEye2 uses a two-stage generation process. Left (SDXL unCLIP): In the initial stage, latent features are fed into a fine-tuned SDXL \citep{podell_sdxl_2023} with unCLIP technique \citep{ramesh_hierarchical_2022, scotti_mindeye2_2024} to generate images, which closely match the originals. Right (MindEye2):  These initial images are then refined in a second stage using base SDXL and captions predicted from the latent features (via GiT Image2Text modules; \citealp{wang2022git}). Because this second stage incorporates text information, the final MindEye2 outputs show lower fidelity than the SDXL unCLIP results.}
  }

\label{Figure_A8}
\end{figure}

\clearpage
\begin{figure}[!t]
  \centering
    \includegraphics[width=1.0\linewidth]{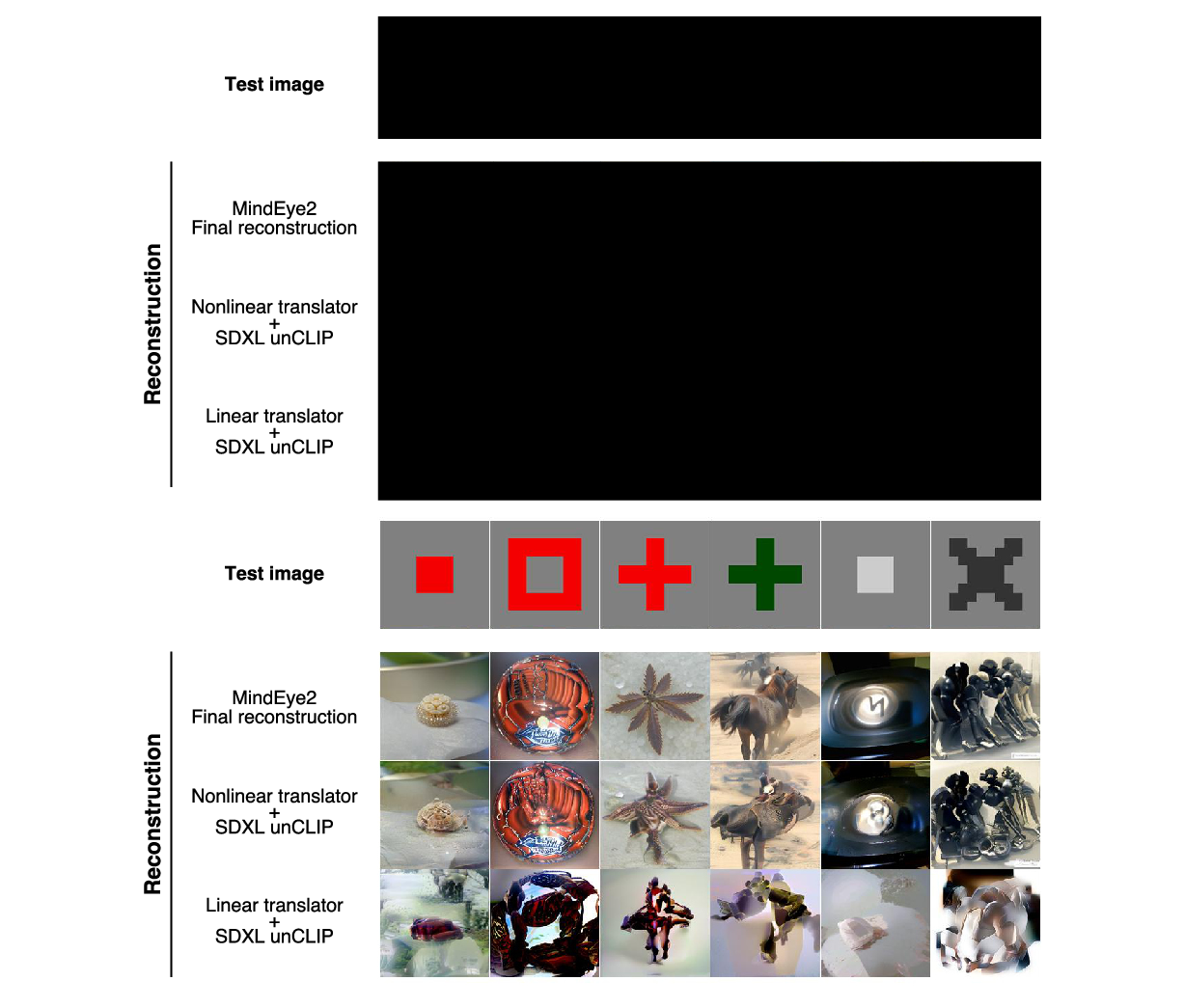}

  \caption{\textbf{Impact of the nonlinear translator on MindEye2 reconstructions.} \rev{The first row shows the test images (Deeprecon), and the second row presents the full MindEye2 reconstructions, which use a nonlinear translator followed by a two-stage generator (SDXL unCLIP and then SDXL with predicted captions). To isolate the effects of the generator utilizing GiT Image2Text modules, the third and fourth rows compare only the first-stage outputs (SDXL unCLIP) using a nonlinear translator (third row) versus a linear translator (fourth row). Nonlinear translators often revert to object categories seen in the training set, whereas linear translators do not exhibit this tendency, suggesting that nonlinear models may be more prone to output dimension collapse.}
  }

\label{Figure_A9}
\end{figure}

\clearpage
\thispagestyle{empty}
\begin{figure}[!t]
  \vspace{-2cm}
  \centering

    \includegraphics[width=1.0\linewidth]{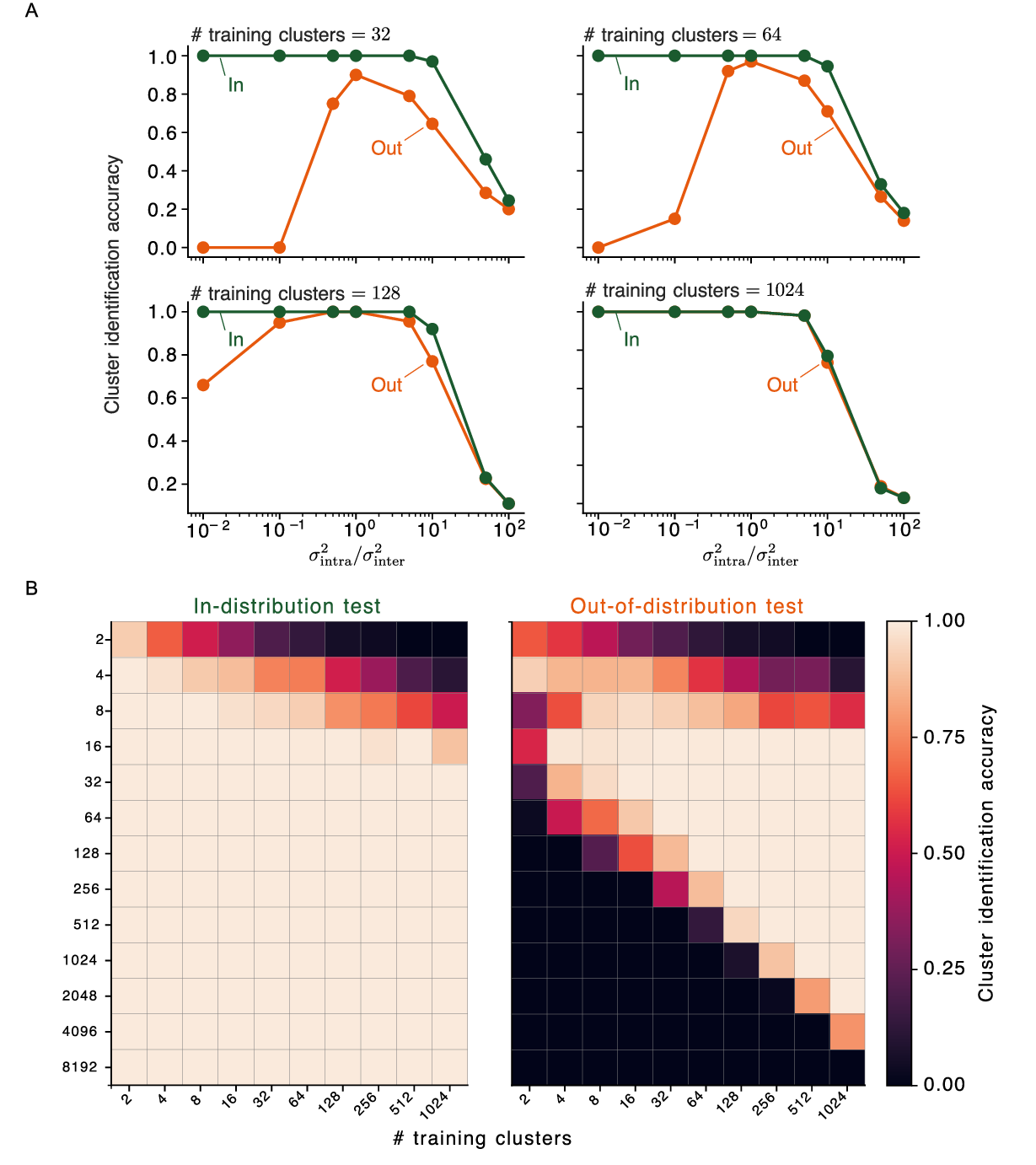}
  \caption{\textbf{Extended simulation results for clustered data analysis.} (\textbf{A}) Relationship between cluster identification accuracy and cluster variance ratio. Each subplot represents a different number of clusters used in model training. 
The $x$-axis represents the cluster variance ratio $\sigma_{\mathrm {intra}}^2/\sigma_{\mathrm {inter}}^2$, 
and the $y$-axis represents the cluster identification performance. The green and orange lines indicate in-distribution and out-of-distribution (OOD) test samples, respectively. Higher variance ratios increase cluster overlap, making in-distribution identification harder. However, they also expand feature space, improving OOD prediction. OOD performance peaks at intermediate variance ratios, where cluster separability and feature space coverage are balanced. (\textbf{B}) Heatmap of cluster 
identification accuracy. These matrices visualize how accuracy changes with 
varying numbers of training clusters and feature space dimensions. The cluster 
variance ratio ($\sigma_{\mathrm {intra}}^2/\sigma_{\mathrm{inter}}^2$) is fixed at $0.1$ for all simulations. 
The left matrix represents in-distribution samples, and the right represents OOD samples. Each cell $(i, j)$ indicates the accuracy for the feature 
dimension $i$ and $j$ training clusters. In the OOD condition, higher feature dimensionality requires more training clusters for robust zero-shot prediction and this relationship is linear, not exponential.
}

\label{Figure_A10}
\end{figure}

\clearpage
\begin{figure}[!t]
  \vspace{-2cm}
  \centering
    \includegraphics[width=1.0\linewidth]{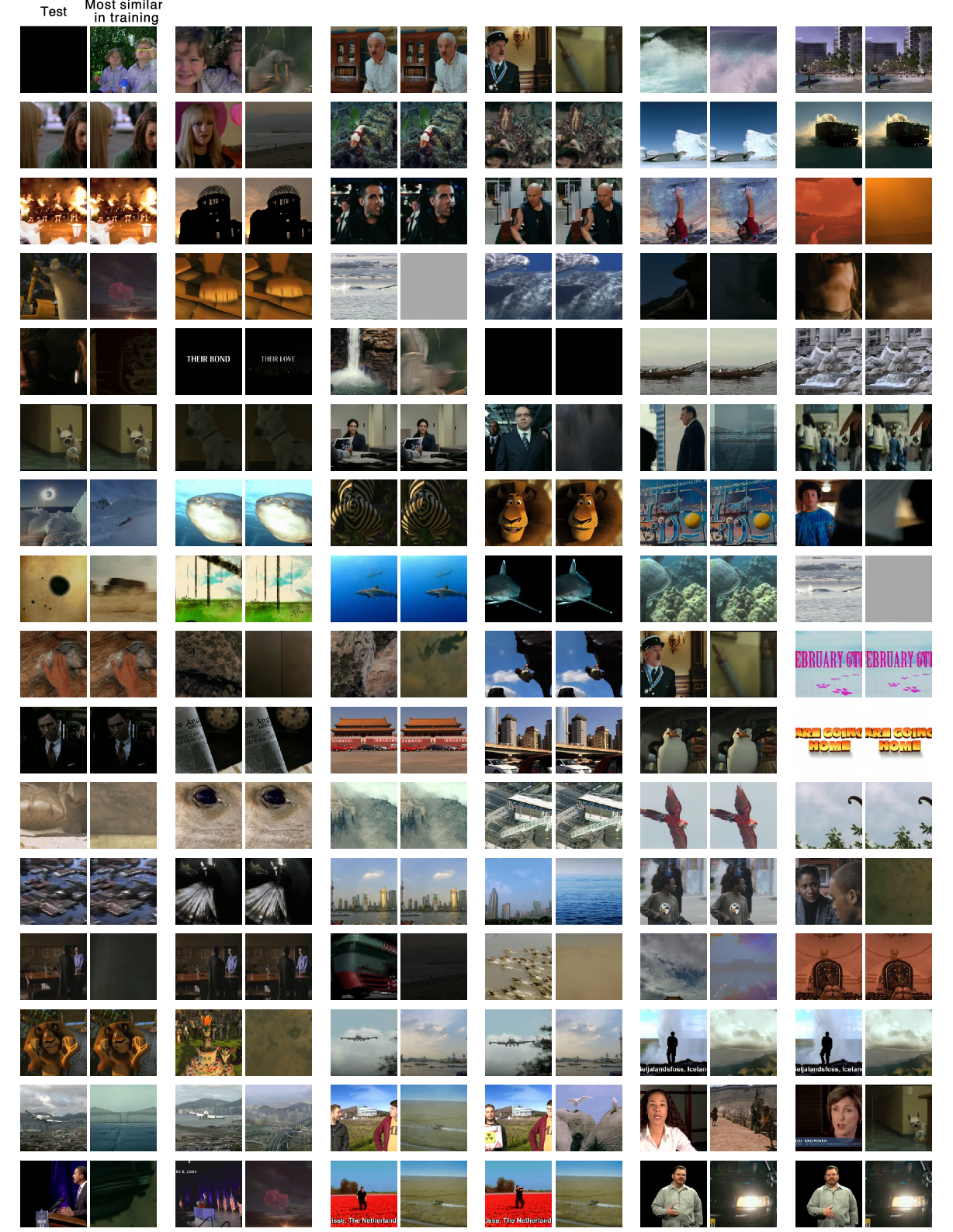}

  \caption{\textbf{Frame similarity analysis in the dataset of \citep{nishimoto_reconstructing_2011}.} Movie scenes were detected from the movie stimuli. Then, for the first and last frames of each test scene, we identified the frames from the training movies with the closest Euclidean distance. Our analysis revealed that 37 out of the 48 scenes in the test set contained frames that were nearly identical to those in the training set. 
}

\label{Figure_A11}
\end{figure}


\end{document}